\documentclass[a4paper,11pt]{article}

\usepackage{graphicx}
\usepackage{amsmath}
\usepackage{a4wide}
\usepackage{amsfonts}
\usepackage{url}
\usepackage{multirow}

\newenvironment{pf}{\textbf{Proof:}}{\hspace{\stretch{1}}$\square$}

\newtheorem{thm}{Theorem}
\newtheorem{df}{Definition}
\newtheorem{pr}{Proposition}
\newtheorem{dthm}{Definition-Theorem}
\newtheorem{lem}{Lemma}

\newtheorem{rk}{Remark}

\DeclareMathOperator{\trace}{Tr}
\DeclareMathOperator{\I}{I}
\newcommand{\ketbra}[2]{| #1 \rangle \langle #2 |}
\renewcommand{\P}{\mathbb{P}}
\newcommand{\E}{\mathbb{E}}
\newcommand{\N}{\mathbb{N}}
\newcommand{\C}{\mathbb{C}}
\newcommand{\M}{\mathcal{M}}

\newcommand{\isom}{\simeq}

\title{Quantum Trajectories in Random Environment: the Statistical Model for a Heat Bath}
\author{Ion Nechita\footnote{D\'epartement de Math\'ematique et Statistique, Universit\'e d'Ottawa, 585 King Edward, Ottawa, ON, K1N6N5 Canada and CNRS, Institut Camille Jordan Universit\'e  Lyon 1, 43 Bd du 11 Novembre 1918, 69622 Villeurbanne, 
France. Email : inechita@uottawa.ca}
$\;$ and Cl\'ement Pellegrini\footnote{University of KwaZulu Natal, NITHeP, Private Bag X54001, Durban 4000, South Africa. Email : pelleg@math.univ-lyon1.fr}}

\begin{document}

\maketitle

\begin{abstract}
In this article, we derive the stochastic master equations corresponding to the statistical model of a heat bath. These stochastic differential equations are obtained as continuous time limits of discrete models of quantum repeated measurements. Physically, they describe the evolution of a small system in contact with a heat bath undergoing continuous measurement. The equations obtained in the present work are qualitatively different from the ones derived in \cite{A1P1}, where the Gibbs model of heat bath has been studied. It is shown that the statistical model of a heat bath provides clear physical interpretation in terms of emissions and absorptions of photons. Our approach yields models of random environment and unravelings of stochastic master equations. The equations are rigorously obtained as solutions of martingale problems using the convergence of Markov generators.   
\end{abstract}

\section*{Introduction}

The theory of \emph{Quantum Trajectories} consists in studying the evolution of the state of an open quantum system undergoing continuous indirect measurement. The most basic physical setting consists of a small system, which is the open system, in contact with an environment. Usually, in quantum optics and quantum communication, the measurement is indirectly performed on the environment \cite{BAR,Book,B5,B2,F1,W1,W2}. In this framework, the reduced time evolution of the small system, obtained by tracing over the degrees of freedom of the environment, is described by stochastic differential equations called \emph{stochastic Schr\"odinger equations} or \emph{stochastic Master equations}. The solutions of these equations are called \emph{Continuous Quantum Trajectories}. In the literature, two generic types of equations are usually considered
\begin{enumerate}
\item \emph{Diffusive} equations
\begin{equation}\label{D} d\rho_t=\mathcal{L}(\rho_t)dt+\Big(C\rho_t+\rho_tC^\star-\trace\Big[\rho_t(C+C^\star)\Big]\,\rho_t\Big)dW_t,
\end{equation}
where $(W_t)_{t \geq 0}$ is a one dimensional Brownian motion.
\item \emph{Jump} equations
\begin{equation}\label{J}
d\rho_t=\mathcal{L}(\rho_t)dt+\left(\frac{C\,\rho_t\,C^\star}{\trace\big[C\,\rho_t\,C^\star\big]}-\rho_t\right)\Big(d\tilde{N}_t-\trace\big[C\,\rho_t\,C^\star\big]dt\Big),
\end{equation}
where $(\tilde{N}_t)_{t \geq 0}$ is a counting process with stochastic intensity $t \mapsto \int_0^t\trace\big[C\,\rho_s\,C^\star\big]ds.$
\end{enumerate}

Physically, equation (\ref{D}) describes photon detection models called \emph{heterodyne} or \emph{homodyne} detection \cite{BAR,Book,W1,W2}. The equation (\ref{J}) relates \emph{direct} photon detection model \cite{BAR,W1,W2}. The driving noise depends then on the type of measurement. Mathematically, a rigorous approach for justifying these equations is based on the theory of \emph{Quantum Stochastic Calculus} \cite{B1, B7, Bout1, Part1}. In such a physical setup, the action of the environment (described usually by a Fock space) on the small system is modeled by \emph{quantum noises} \cite{A1, A2, gardiner}. The evolution is then described by the so-called \emph{Quantum Stochastic Differential Equations} \cite{A1, A2,Part1, Fagn1}. Next, by using the \emph{quantum filtering} \cite{Bel1, Bout2, Bout3} technique, one can derive the stochastic Schr\"odinger equations by taking into account the indirect observations. Another approach, not directly connected with quantum stochastic calculus, consists in using \emph{instrumental operator process} and notion of \emph{a posteriori} state \cite{BAR,Book,B2, B3, B4,infinite}.

In this work, we shall use a different approach, introduced recently by the second author in \cite{pelleg1, pelleg2, pelleg3}. This discrete-time model of indirect measurement, called \emph{Quantum Repeated Measurements} is based on the model of \emph{Quantum Repeated Interactions} \cite{AJ, AP, AP2} introduced by S. Attal and Y. Pautrat. The setup is the following: a small system $\mathcal{H}$ is in contact with an infinite chain, $\bigotimes_{k=1}^\infty\mathcal{E}_k$, of identical and independent quantum systems, that is $\mathcal{E}_k=\mathcal{E}$ for all $k$. The elements of the chain interact with the small system, one after the other, each interaction having a duration $\tau > 0$. After each interaction, a quantum measurement is performed on the element of the chain that has just been in contact with the small system. Each measurement involves a random perturbation of the state of the small system, the randomness being given by the outcome of the corresponding quantum measurement. The complete evolution of the state of the small system is described by a Markov chain depending on the time parameter $\tau$. This Markov chain is called a \emph{Discrete Quantum Trajectory}. By rescaling the intensity of the interaction between the small system and the elements of the chain in terms of $\tau$, it has been shown in \cite{pelleg1, pelleg2} that the solutions of equations (\ref{D},\ref{J}) can be obtained as limits of the discrete quantum trajectories when the time step $\tau$ goes to zero. 

In \cite{pelleg1, pelleg2}, the author investigated the case when the reference state of each element of the chain is the ground state (this corresponds also to models at zero temperature). This setup was generalized in \cite{A1P1}, where Gibbs states with positive temperature were considered and the corresponding equations were derived. In the present work, we go beyond this generalization and study the \emph{statistical model for the temperature state} of the chain. More precisely, the initial state of the elements of the chain is a statistical mixture of ground and excited states. It is important to notice that both the Gibbs model as well as the ground state model are deterministic. Let us stress that, in the case where no measurement is performed after each interaction, both the Gibbs and the statistical model give rise to the same deterministic limit evolution. This limit behavior confirms the idea that a mixed quantum state and a probabilistic mixture of pure states represent the same physical reality. Quite surprisingly, we show that, when adding measurement, the limit stochastic differential equations are of different nature: for the Gibbs model the only possible limit evolutions are deterministic or diffusive, whereas for the statistical model jump evolutions becomes a possibility. Furthermore, the Gibbs model limit equations involve at most one random noise, whereas two driving noises may appear at the limit when considering in the statistical model.

The article is structured as follows. In Section \ref{sec:rep_int}, we introduce the different discrete models of quantum repeated interactions and measurements. In our approach, we present the statistical model of the thermal state as the result of a quantum measurement applied to each element of the chain \emph{before} each interaction. Next, we describe the random evolution of the open system by deriving discrete stochastic equations. In Section \ref{sec:cont_models}, we investigate the continuous time models obtained as limits of the discrete models when the time-step parameter goes to zero. We remind the results of \cite{A1P1} related to the thermal Gibbs model and we describe the new continuous models related to the thermal statistical model. Section \ref{sec:discussion} is devoted to the analysis of the different models. The qualitative differences between the continuous time evolutions are illustrated by concrete examples. Within these examples, it is shown that the statistical approach provides clear physical interpretations which cannot be reach when considering the Gibbs model. We show that model at zero temperature (each element of the chain is at the ground state) can be recovered by the statistical model; however, this is not possible with the Gibbs model. Moreover, we show that considering the statistical model allows to obtain \emph{unravelings} of heat master equations with a measurement interpretation. Section \ref{sec:proofs} contains the proofs of the convergence of the discrete time model to the continuous model. Such results are based on Markov chain approximation techniques using the notion of convergence of Markov generators and martingale problems.

\section{Quantum Repeated Interactions and Discrete Quantum Trajectories}\label{sec:rep_int}

In this section we present the mathematical model of quantum repeated measurements. In the first subsection we briefly recall the model of quantum repeated interactions \cite{AP} and in the second subsection we describe three different situations of indirect quantum measurements, in which environment particles are measured before and/or after each interaction. Discrete evolution equations are obtained in each case.

\subsection{Quantum Repeated Interactions Model without Measurement}

Let us introduce here the mathematical framework of quantum repeated interactions. We consider a small system  $\mathcal{H}$ in contact with an infinite chain of identical and independent quantum systems. Each piece of the chain is represented by a Hilbert space $\mathcal{E}$. Each copy of $\mathcal{E}$ interacts, one after the
other, with the small system $\mathcal{H}$ during a time $\tau$. Note that all the Hilbert spaces we consider are complex and finite dimensional.

We start with the simpler task of describing a single interaction between the small system $\mathcal{H}$ and one piece of the environment $\mathcal E$. Let $\rho$ denote the state of $\mathcal{H}$ and let $\sigma$ be the state of $\mathcal{E}$. States are a positive self-adjoint operators of trace one; in Quantum Information Theory they are also called density matrices. The coupled system is described by the tensor product $\mathcal{H}\otimes\mathcal{E}$ and the initial state is in a product form $\rho \otimes \sigma$. The evolution of the coupled system is given by a total Hamiltonian acting on 
$\mathcal{H}\otimes\mathcal{E}$ 
\[H_{\text{tot}}=H_0\otimes I+I\otimes H+H_{\text{int}},\]
where the operators $H_0$ and $H$ are the free Hamiltonians of the systems $\mathcal{H}$ and $\mathcal{E}$ respectively, and the operator $H_{\text{int}}$ is the interaction Hamiltonian. The operator $H_{\text{tot}}$ gives rise to an unitary operator
\[U=\exp(-i\tau H_{\text{tot}}),\]
where $\tau$ represents the time of interaction. After the
interaction, in the Schr\"odinger picture, the final state of the coupled system is
\[\mu=U(\rho\otimes\sigma)U^\star.\]

In order to describe \emph{all} the repeated interactions, we need to describe an infinite number of quantum systems. The Hilbert space of all possible states is given by the countable tensor product
\[\mathbf{\Gamma}=\mathcal{H}\otimes\bigotimes_{k=1}^\infty\mathcal{E}_k = \mathcal H \otimes \mathbf\Phi,\]
where $\mathcal{E}_k\isom\mathcal{E}$ for all $k \geq 1$. If
$\{e_0,e_1,\ldots,e_K\}$ denotes an orthonormal basis of
$\mathcal{E}\isom\mathbb{C}^{K+1}$, the orthonormal basis of $\mathbf\Phi=\bigotimes_{k=1}^\infty\mathcal{E}_k$ is constructed with respect to the stabilizing sequence $e_0^{\otimes \N^*}$ (we shall not develop the explicit construction of the countable tensor product since we do not need it in the rest of the paper; we refer the interested reader to \cite{AP} for the complete details). 

Let us now describe the interaction between $\mathcal H$ and the $k$-th piece of environment $\mathcal E_k$, from the point of view of the global Hilbert space $\mathbf\Gamma$. The quantum interaction is given by an unitary operator $U_k$ which acts like the operator $U$ on the tensor product $\mathcal{H} \otimes \mathcal{E}_k$ and like the identity operator on the rest of the space $\mathbf\Gamma$. In the Schr\"odinger picture, a state $\eta$ of $\mathbf{\Gamma}$ evolves as a closed system, by unitary conjugation 
$$\eta\longmapsto U_k\,\eta\,U_k^\star.$$
Therefore, the whole procedure up to time $k$ can be described by an unitary operator $V_k$ defined recursively by
\begin{equation}\left\{\begin{array}{lcl} V_{k+1}&=&U_{k+1}V_k \\
  V_0&=&I\end{array}\right.\end{equation}

In more concrete terms we consider the initial state $\mu=\rho\otimes\bigotimes_{k=1}^\infty\sigma$ for the small system coupled with the chain (notice that all the elements of the chain are initially in the same state $\sigma_k = \sigma$). After $k$ interactions, the
reference state is given by
$$\mu_k=V_k\,\mu\,V_k^\star.$$

Since we are interested only in the evolution of the small system $\mathcal H$, we discard the environment $\mathbf\Phi$. The reduced dynamics of the small system is then given by the partial trace on the degrees of freedom of the environment. If $\alpha$ denotes a state on $\mathbf\Gamma$, we denote by $\trace_{\mathbf\Phi}[\alpha]$ the partial trace of $\alpha$ on $\mathcal{H}$ with respect to the environment space $\mathbf\Phi = \bigotimes_{k=1}^\infty\mathcal{E}_k$. 
We recall the definition of the partial trace operation.

\begin{dthm}
Let $\mathcal{H}$ and $\mathcal{K}$ be two Hilbert spaces. For
all state $\alpha$ on $\mathcal{H}\otimes\mathcal{K}$, there
exists a unique state on $\mathcal{H}$ denoted by
$\trace_{\mathcal{K}}[\alpha]$ which satisfies
\[\trace\big[\trace_{\mathcal{K}}[\alpha]X\big]=\trace[\alpha(X\otimes \I_{\mathcal{K}})],\]
for all $X\in\mathcal{B}(\mathcal{H})$. The state
$\trace_{\mathcal{K}}[\alpha]$ is called the partial trace of
$\alpha$ on $\mathcal{H}$ with respect to $\mathcal{K}$.
\end{dthm}
With this notation, the evolution of the state of the small system is given by
\begin{equation}\label{reduced}
\rho_k=\trace_{\mathbf\Phi}\big[\,\mu_k\,\big].
\end{equation}
The reduced dynamics of $(\rho_k)$ is entirely described by the following proposition \cite{AP,NP}.
\begin{pr}\label{reducedd}
The sequence of states $(\rho_k)_k$ defined in equation (\ref{reduced}) satisfies the recurrence relation
$$\rho_{k+1}=\trace_{\mathcal{E}}\big[\,U(\rho_k\otimes\sigma)U^\star\,\big].$$
Furthermore, the application
\begin{align*}
L:\mathcal B(\mathcal H) &\to \mathcal B(\mathcal H) \\
 X &\mapsto \trace_{\mathcal{E}}[U(X\otimes\sigma)U^\star]
\end{align*}
defines a trace preserving completely positive map (or a \emph{quantum channel}) and the state of the small system after $k$ interactions is given by
\begin{equation}\label{mmaster}
 \rho_k = {L}^k(\rho_0).
\end{equation}
\end{pr}

\subsection{Quantum Repeated Interactions with Measurement}
In this section we introduce Quantum Measurement in the model of quantum repeated interactions and we show how equation (\ref{mmaster}) is modified by the different observations. We shall study three different situations of indirect measurement, as follows:

\begin{enumerate}
 \item The first model concerns ``quantum repeated measurements" \textbf{before} each interaction. It means that we perform a measurement of an observable on each copy of $\mathcal{E}$ before the interaction with $\mathcal{H}$. We call such a setup ``Random Environment" (we shall explain the terminology choice later on).
 \item The second model concerns "quantum repeated measurements" \textbf{after} each interaction. It means that we perform a measurement of an observable on each copy of $\mathcal{E}$ after the interaction with $\mathcal{H}$. We call such a setup ``Usual Indirect Quantum Measurement".
 \item The third setup is a combination of the two previous models. Two quantum measurements (of possibly different observables) are performed on each copy of $\mathcal{E}$, one \textbf{before} and one \textbf{after} each interaction with $\mathcal{H}$. Such a setup is called ``Indirect Quantum Measurement in Random Environment"
\end{enumerate}

In all the cases, the measurement is called indirect because the small system is not directly observed, the measurement being performed on an auxiliary system (an element of the chain) which interacted previously with the system. The main purpose of this work is to study and analyze the three different limit behaviors obtained when the interaction time $\tau$ goes to zero (see Section 2). Let us mention that the second setup has been studied in detail in \cite{pelleg1, pelleg2, pelleg3}. We chose to describe in great detail the more general case of the third model, since the other two models can be easily recovered from the third one, by choosing to measure the trivial observable $\I$.

\subsubsection{Indirect Quantum Measurement in Random Environment}\label{QMRE}

In order to make the computations more easy to follow, we shall focus on the case where the environment is a chain of qubits (two-dimensional quantum systems). Mathematically, this is to say that $\mathcal{E}=\mathbb{C}^2$. 

Let us start by making more precise the physical model for one copy of $\mathcal{E}$. To this end, we consider $\{e_0,e_1\}$ an orthonormal basis of $\mathcal{E}$, which diagonalizes the Hamiltonian 
\[H = \begin{pmatrix}
\gamma_0&0\\
0&\gamma_1
\end{pmatrix},\]
where we suppose that $\gamma_0<\gamma_1$. The reference state $\sigma$ of the environment corresponds to a Gibbs thermal state at positive temperature, that is
\begin{equation}\label{Gibbsstate}\sigma=\frac{e^{-{\beta}H}}{\trace\left[e^{-{\beta}H}\right]},\,\,\,\,\textrm{with}\,\,\,\,{\beta}=\frac{1}{KT},\end{equation}where $T$ corresponds to a finite strictly positive temperature and $K$ is a constant. In the basis $\{e_0, e_1\}$, $\sigma$ is diagonal
$$\sigma=p\vert e_0\rangle\langle e_0\vert+(1-p)\vert e_1\rangle\langle e_1\vert,$$
with
\[p=\frac{e^{-\beta\gamma_0}}{e^{-\beta\gamma_0}+e^{-\beta\gamma_1}}.\]
Notice that since $\beta>0$, we have $0<p<1$.

We are now in position to describe the measurement before the interaction. We consider a diagonal observable $A$ of $\mathcal{E}$ of the form
$$A=\lambda_0 \ketbra{e_0}{e_0} + \lambda_1 \ketbra{e_1}{e_1}.$$
The extension of the observable $A$ to an observable of $\mathcal{H}\otimes\mathcal{E}$ is $\I\otimes A$. According to the axioms of Quantum Mechanics, the outcome of the measurement of the observable $\I\otimes A$ is an element of its spectrum, the result being random. If the initial state (before the interaction) is $\rho\otimes\sigma$, we shall observe the eigenvalue $\lambda_i$ with probability
\[\P[\lambda_i \text{ is observed}]=\trace\big[(\rho\otimes\sigma)\,\,\I\otimes P_i\big] = \trace[\sigma P_i],\quad i=0,1\]
where $P_i = \ketbra{e_i}{e_i}$ are the eigenprojectors of $A$. It is straightforward to see that in this case
\[\P[\lambda_0\text{ is observed}]=p=1-\P[\lambda_1 \text{ is observed}].\]
Furthermore, according to the wave packet reduction principle, if the eigenvalue $\lambda_i$ is observed, the initial state $\rho\otimes\sigma$ is modified and becomes
\begin{equation}\label{mu}\mu^1_i=\frac{\I\otimes P_i\,\,(\rho\otimes\sigma)\,\, \I\otimes P_i}{\trace\big[(\rho\otimes\sigma)\,\, \I\otimes P_i\big]}= \rho \otimes \frac{P_i \sigma P_i}{\trace[\sigma P_i]}.\end{equation}
This defines naturally a random variable $\mu^1$ valued in the set
of states on $\mathcal{H}\otimes\mathcal{E}$. More precisely, the state
$\mu^1$ takes the value $\mu^1_0=\rho\otimes\vert e_0\rangle\langle e_0\vert$ with probability
$\trace\big[(\rho\otimes\sigma)\,\,\vert e_0\rangle\langle e_0\vert\big]=p$ and the
value $\mu^1_1=\rho\otimes\vert e_1\rangle\langle e_1\vert$ with probability $1-p$.\\
\begin{rk} Since both the initial state of the system and the observable measured have product form, only the state of $\mathcal{E}$ is modified by the measurement before the interaction. Instead of describing the evolution of the coupled system, we could have considered that the state of $\mathcal{E}$ is a random variable $\sigma_i^1$ where $\sigma_i^1$ is either $\ketbra{e_0}{e_0}$ with probability $p$ either $\ketbra{e_1}{e_1}$ with probability $1-p$. This is the statistical model for a thermal state and its random character justifies the name ``Random environment". In conclusion, we could have replaced from the start the setup \emph{(Gibbs state + Quantum measurement)} with the probabilistic setup \emph{Random environment}, the results being identical. We shall give more details and comments on this point of view in the Section \ref{sec:discussion}.
\end{rk}

We now move on to describe the second measurement, which is performed after the interaction. In this case we consider an arbitrary (not necessarily diagonal in the basis $\{e_0, e_1\}$) observable $B$ of $\mathcal{E}$ which admits a spectral decomposition
\[B=\alpha_0 Q_0+\alpha_1 Q_1,\]
where $Q_j$ corresponds to the eigenprojector associated with the eigenvalue $\alpha_j$. Let $\mu^1$ be the random state after the first measurement. After the interaction, the state on $\mathcal{H}\otimes\mathcal{E}$ is
\[\eta^1_i=U\,\mu^1_i\,U^\star,\quad i=0,1.\]
Now, assuming that the measurement of the observable $A$ (before the interaction) has given the result $\lambda_i$, the probability of observing the eigenvalue $\alpha_j$ of $B$ is given by
\[\P[\alpha_j\text{ is observed}]=\trace\left[\eta^1_i\,\,\I\otimes Q_j\right].\]
and the state after the measurement becomes
\[\theta^1_{i,j}=\frac{\I\otimes Q_j\,\,\eta_i^1\,\, \I\otimes Q_j}{\trace\left[\eta^1_i\,I\otimes Q_j\right]}.\]
The random state $\theta^1$ (which takes one of the values $\theta^1_{i,j}$) on $\mathcal{H}\otimes\mathcal{E}$ describes the random result of the two indirect measurements which were performed before and after the interaction.
\bigskip

Having described the interaction between the small system $\mathcal H$ and one copy of $\mathcal E$, we look now at the repeated procedure on the whole system $\mathbf\Gamma$. The probability space underlying the outcomes of the repeated quantum measurements before and after each interaction is given by $\Omega=(\Sigma_A\times\Sigma_B)^{\mathbb{N}^\star}$, where $\Sigma_A=\{0,1\}$ corresponds to the index of the eigenvalues of the observable $A$ and $\Sigma_B=\{0,1\}$ for the ones of $B$. On $\Omega$, we consider the usual cylinder $\sigma$-algebra $\Lambda$ generated by the cylinder sets
$$\Lambda_{(i_1,j_1),\ldots,(i_k,j_k)}=\{(\omega,\varphi)\in
(\Sigma_A\times\Sigma_B)^{\mathbb{N}^\star}\,|\,\omega_1=i_1,\ldots,\omega_k=i_k
,\varphi_1=j_1,\ldots,\varphi_k=j_k\}.$$

Now, we shall define a probability measure describing the results of the repeated quantum measurements. To this end, we introduce the following notation. For an operator $Z$ on $\mathcal{E}_j$, we note $Z^{(j)}$ the extension of $Z$ as an operator on $\mathbf\Gamma$, which acts as $Z$ on the $j$-th copy of $\mathcal E$ and as the identity on $\mathcal H$ and on the other copies of $\mathcal E$:
$$Z^{(j)}=\I\otimes\bigotimes_{p=1}^{j-1}\I\otimes Z\otimes\bigotimes_{p\geq
j+1}\I.$$
Furthermore, for all $k\geq 1$ and $\{(i_1,j_1),\ldots,(i_k,j_k)\}\in(\Sigma_A\times\Sigma_B)^k$, we put
\begin{equation}\label{nonormalised}
\tilde{\mu}_k\big((i_1,j_1),\ldots,(i_k,j_k)\big)=\left(\prod_{s=1}^k Q^{(s)}_{j_s}\right)\,V_k\,\left(\prod_{s=1}^k P^{(s)}_{i_s}\right)\,
\mu\,\left(\prod_{s=1}^k P^{(s)}_{i_s}\right)\,V_k^\star\,
\left(\prod_{s=1}^k Q^{(s)}_{j_s}\right),
\end{equation}
where $P_i$ and $Q_j$ are the respective eigenprojectors of $A$ and $B$ and $\mu=\rho\otimes\bigotimes_{k=1}^{\infty}\sigma_k$, with $\sigma_k=\sigma = p \ketbra{e_0}{e_0}+(1-p) \ketbra{e_1}{e_1}$ for all $k\in\mathbb{N}^\star$, is the initial state on $\mathbf\Gamma$. Notice that the products in the previous equation need not to be ordered, since two operators $X^{(i)}$ and $Y^{(j)}$ commute whenever $i \neq j$. In the same vein, the following important commutation relation
\[Q^{(k)}_{i_k}U_kP^{(k)}_{i_k}\ldots Q^{(1)}_{i_1}U_1P^{(1)}_{i_1}=\left(\prod_{s=1}^k Q^{(s)}_{j_s}\right) \,V_k\, \left(\prod_{s=1}^k P^{(s)}_{i_s}\right),\]
shows that the operator $\tilde{\mu}_k\big((i_1,j_1),\ldots,(i_k,j_k)\big)$ in Eq. (\ref{nonormalised}) is actually the non normalized state of the global system after the observation of eigenvalues $\lambda_{i_1},\ldots,\lambda_{i_k}$ for $k$ first measurements of $A$ and $\alpha_{j_1},\ldots,\alpha_{j_k}$ for the $k$ first measurements of the observable $B$.

We have now all the elements needed to define a probability measure on the cylinder algebra $\Lambda$ by
\[\P[\Lambda_{(i_1,j_1),\ldots,(i_k,j_k)}]=\trace[\tilde{\mu}_k\big((i_1,j_1),\ldots,(i_k,j_k)\big)].\]
This probability measure satisfies the Kolmogorov Consistency Criterion, hence we can extend it to the whole $\sigma$-algebra $\Lambda$ to the unique probability measure $\P$ with these finite dimensional marginals.

The global random evolution on $\mathbf\Gamma$ is then described by the random sequence $(\tilde{\rho}^k)$
\[\begin{array}{crcl}\tilde{\rho}_k: & \Omega
  &\longrightarrow&
\mathcal{B}(\mathbf{\Gamma})\\
 & (\omega,\varphi) & \longmapsto & \tilde{\rho}_k(\omega,\varphi)=
 \displaystyle{\frac{\tilde{\mu}((\omega_1,\varphi_1),\ldots,
(\omega_k,\varphi_k))}{\trace[\tilde{\mu}_k((\omega_1,\varphi_1),\ldots,
(\omega_k,\varphi_k))]}}
\end{array}\]
This random sequence describes the random modification involved by
the result of measurement before and after the interactions. In
order to recover the measurement setup only before or only after
the interactions, one has just to delete the projector $P^{(j)}_{i_j}$ or
$Q^{(j)}_{i_j}$ in equation (\ref{nonormalised}).

The reduced evolution of the small system is obtained by the partial trace operation:
\begin{equation}
\label{reduceddd}
\rho_k\big(\omega,\varphi\big)=\trace_{\mathbf\Phi}\left[\tilde{\rho}_k\big(\omega,\varphi\big)\right]
\end{equation}
for all
$(\omega,\varphi)\in\Omega$
and all $k\in\mathbb{N}^\star$. The random sequence $(\rho_k)_{k \geq 1}$ is called a \emph{Discrete Quantum Trajectory}. It describes the random modification of the small system
undergoing the sequence of successive measurements.

\begin{rk}
The dynamics of the sequence of states $\rho_k$ can be seen as a \emph{random walk in random environment} dynamics in the following way. Assume that all the elements of the chain are measured before the first interaction; the results of this procedure define a random environment in which the small system will evolve. All the randomness coming from the measurement before each interaction is now contained in the environment $\omega$. Given a fixed value of the environment $\omega$, the small system interacts repeatedly with the chain (whose states depend on $\omega$) and the random results of the repeated measurement of the second observable $B$ are encoded in $\varphi$. In this way, the global evolution of $\rho$ can be seen as a random walk (where random modifications of the states are due to the second measurement) in a random environment (generated by the measurements before each interaction). 
\end{rk}

\subsubsection{Discrete Evolution Equations}\label{DEV}

In this section, using the Markov property of the discrete quantum trajectories, we obtain discrete evolution equations
which are random perturbation of the Master equation (\ref{mmaster}) given in
Proposition \ref{reducedd}. The Markov property of the random sequence $(\rho_k)_k$ is expressed as follows.

\begin{pr}\label{mmaa}
The random sequence of states $(\rho_k)_k$ on $\mathcal{H}$ defined by the formula $(\ref{reduceddd})$ is a Markov chain on $(\Omega,\Lambda,\P)$. More precisely, we have the following random evolution equation
 \begin{equation}\label{eq}
\rho_{k+1}(\omega,\varphi)=\sum_{i,j\in\{0,1\}}\frac{\mathcal G_{ij}(\rho_k(\omega, \varphi))}{\trace\big[\mathcal G_{ij}(\rho_k(\omega, \varphi))\big]}\,\mathbf{1}^{k+1}_{ij}(\omega,\varphi),
\end{equation}
where 
\[\mathcal G_{ij}(\rho) = \trace_{\mathcal E}\left[(I\otimes
Q_j)\,\,U\,\left(\I\otimes P_i
\,\left(\rho\otimes\sigma\right)\,\I\otimes
P_i\right)\,U^\star\,\,(\I\otimes Q_j)\right] \]
and $\mathbf{1}^{k+1}_{ij}(\omega,\varphi)=\mathbf{1}_{ij}(\omega_{k+1},\varphi_{k+1}) = \mathbf{1}_{i}(\omega_{k+1})\mathbf{1}_{j}(\varphi_{k+1})$ for all $(\omega,\varphi)\in (\Sigma_A\times\Sigma_B)^{\mathbb{N}^\star}.$
\end{pr}

The equation (\ref{eq}) is called a \textit{Discrete Stochastic
Master Equation}. In order to make more explicit the equation (\ref{eq}) and to compute the partial trace, we introduce a suitable basis for $\mathcal{H}\otimes\mathcal{E}$, which is
$\{e_0\otimes e_0,e_1\otimes e_0,e_0\otimes e_1,e_1\otimes
e_1\}.$ In this basis, the unitary operator $U$ can be written in block format in the following way
\[U=\begin{pmatrix}
L_{00}&L_{01}\\L_{10}&L_{11}
\end{pmatrix},\]
where $L_{ij}$ are operators in $\M_2(\C)$. We shall treat two different situations, depending on the form of the observable $B$ that is being measured after each interaction. On one hand we consider the case where the observable $B$
of $\mathcal{E}$ is diagonal in the basis $(e_0,e_1)$ and on the other hand we consider the case where $B$ is non diagonal.

Let us start with the case where the observable $B$ is diagonal in the basis $\{e_0, e_1\}$, that is $B=\alpha_0 \ketbra{e_0}{e_0}+\alpha_1 \ketbra{e_1}{e_1}$. In this case, equation (\ref{eq}) becomes
\begin{eqnarray}\label{tata}
\lefteqn{\rho_{k+1}(\omega,\varphi)=}\nonumber\\&&\frac{L_{00}\,\rho_k(\omega,\varphi)\,L_{00}^\star}
{\trace[L_{00}\,\rho_k(\omega,\varphi)\,L_{00}^\star]}\mathbf{1}_{00}(\omega_{k+1},\varphi_{k+1})+
\frac{L_{10}\,\rho_k(\omega,\varphi)\,L_{10}^\star}
{\trace[L_{10}\,\rho_k(\omega,\varphi)\,L_{10}^\star]}\mathbf{1}_{01}(\omega_{k+1},\varphi_{k+1})\nonumber\\&&
+\frac{L_{01}\,\rho_k(\omega,\varphi)\,L_{01}^\star}
{\trace[L_{01}\,\rho_k(\omega,\varphi)\,L_{01}^\star]}\mathbf{1}_{10}(\omega_{k+1},\varphi_{k+1})+
\frac{L_{11}\,\rho_k(\omega,\varphi)\,L_{11}^\star}
{\trace[L_{11}\,\rho_k(\omega,\varphi)\,L_{11}^\star]}\mathbf{1}_{11}(\omega_{k+1},\varphi_{k+1}).\nonumber\\
\end{eqnarray}

Usually, a stochastic Master equation appears as a random perturbation of the Master equation (see equations (\ref{D}, \ref{J}) in the Introduction). Moreover, the noises driving the equations are centered, that is of zero mean (this is the case of the Brownian motion and the counting process compensated with the stochastic intensity in equations (\ref{D}, \ref{J})).  In order to obtain a similar description in the discrete case, we introduce the following random variables
\begin{equation}
 X_k\big(\omega,\varphi\big)=\frac{\mathbf{1}_{10}(\omega_k,\varphi_k)+\mathbf{1}_{11}(\omega_k,\varphi_k)-(1-p)}{\sqrt{p(1-p)}},\quad k\in\N^\star.
 \end{equation}
 
Now, we rewrite equation (\ref{tata}) in terms of the random variables $X_k$, $\mathbf{1}_{01}$ $\mathbf{1}_{10}$:
\begin{eqnarray}\label{tata1}
\lefteqn{\rho_{k+1}=\Big(p(L_{00}\,\rho_k\,L_{00}^\star+L_{10}\,\rho_k\,L_{10}^\star)+(1-p)(L_{01}\,\rho_k\,L_{01}^\star
+L_{11}\,\rho_k\,L_{11}^\star)\Big)\mathbf{1}}\nonumber\\
&&+\Big(-\sqrt{p(1-p)}\left(L_{00}\,\rho_k\,
L_{00}^\star+L_{10}\,\rho_k\,
L_{10}^\star\right)+\sqrt{p(1-p)}\left(L_{11}\,\rho_k\,
L_{11}^\star+L_{01}\,\rho_k\,
L_{01}^\star\right)\Big)X_{k+1}\nonumber\\
&&+\Bigg(-\frac{L_{00}\,\rho_k\,L_{00}^\star}{\trace[L_{00}\,\rho_k\,L_{00}^\star]}+\frac{L_{10}\,\rho_k\,L_{10}^\star}
{\trace[L_{10}\,\rho_k\,L_{10}^\star]}\Bigg)(\mathbf{1}_{01}-p\trace[L_{10}\rho_kL_{10}])\nonumber\\
&&+\Bigg(-\frac{L_{11}\,\rho_k\,L_{11}^\star}{\trace[L_{11}\,\rho_k\,L_{11}^\star]}+\frac{L_{01}\,\rho_k\,L_{01}^\star}
{\trace[L_{10}\,\rho_k\,L_{10}^\star]}\Bigg)(\mathbf{1}_{10}-(1-p)\trace[L_{01}\rho_kL_{01}]).
\end{eqnarray}
It is important to stress out that the last three terms in the previous equation have mean zero: \[\E\big[X_k\big]=\E\big[\mathbf{1}_{01}-p\trace[L_{10}\rho_kL_{10}]\big]=\E\big[\mathbf{1}_{10}-(1-p)\trace[L_{01}\rho_kL_{01}]\big]=0.\]
Moreover, recall that the discrete evolution of Proposition \ref{reducedd}, without measurement, is given by
\begin{equation}\label{r}\rho_{k+1}=L(\rho_k) = \Big(p\big(L_{00}\,\rho_k\,L_{00}^\star+L_{10}\,\rho_k\,L_{10}^\star\big)+(1-p)\big(L_{01}\,\rho_k\,L_{01}^\star
+L_{11}\,\rho_k\,L_{11}^\star\big)\Big).\end{equation}
As a consequence, the discrete stochastic master equation (\ref{tata1}) is written as a perturbation of the discrete Master equation (\ref{r}). 

\begin{rk}In this expression, one can see that the random variable $X_k$ depends only on the outcome of the measurement before the interaction (we sum over the two possible results of the measurement after the interaction). In other words, it means that the random variables $X_k$ describe essentially the perturbation of the measurement before the interaction.  On the other hand, the random variables $\mathbf{1}_{01}$ and $\mathbf{1}_{10}$, conditionally on the result of the first measurement, describe the perturbation involved by the measurement after the interaction. Hence, each term of the equation (\ref{tata1}) that is linked with either $X_k$, $\mathbf{1}_{01}$ or $\mathbf{1}_{10}$ expresses how the deterministic part (\ref{r}) is modified by the results of the different measurements. 
\end{rk}

We now analyze the second case, where the observable $B$ is non-diagonal in the basis $\{e_0,e_1\}$. We write $B=\alpha_0 Q_0+\alpha_1 Q_1$, where the eigenprojectors $Q_i$ are written in the $\{e_0,e_1\}$ basis $Q_i=(q_{kl}^i)_{0\leq k,l\leq1}$. In this case, the operators appearing in equation (\ref{eq}) are given by
\begin{eqnarray*}\mathcal{G}_{0i}(\rho)&=&q^i_{00}L_{00}\rho L_{00}^\star+q^i_{10}L_{00}
\rho L_{10}^\star+q^i_{01}L_{10}\rho L_{00}^\star+q^i_{11}L_{10}\rho L_{10}^\star\nonumber\\
\mathcal{G}_{1i}(\rho)&=&q^i_{00}L_{01}\rho L_{01}^\star+q^i_{10}L_{01}\rho
L_{11}^\star+q^i_{01}L_{11}\rho L_{01}^\star+q^i_{11}L_{11}\rho
L_{11}^\star.
\end{eqnarray*}
As before, in order to obtain the expression of the discrete Master equation as a perturbation of the deterministic Master equation, we introduce the following random variables
\begin{eqnarray}
 X_{k+1}&=&\frac{\mathbf{1}_{10}^{k+1}+\mathbf{1}_{11}^{k+1}-(1-p)}{\sqrt{p(1-p)}}\nonumber\\
Y^0_{k+1}&=&\frac{\mathbf{1}_{01}^{k+1}-p\trace\big[\mathcal{G}_{01}(\rho_k)\big]}{\sqrt{p\trace\big[\mathcal{G}_{01}(\rho_k)\big]\big(1-p\trace\big[\mathcal{G}_{01}(\rho_k)\big]\big)}}\nonumber\\
Y^1_{k+1}&=&\frac{\mathbf{1}_{10}^{k+1}-(1-p)\trace\big[\mathcal{G}_{10}(\rho_k)\big]}{\sqrt{(1-p)\trace\big[\mathcal{G}_{10}(\rho_k)\big]\big(1-(1-p)\trace\big[\mathcal{G}_{10}(\rho_k)\big]\big)}}.
\end{eqnarray}
In terms of these random variables, we get
 \begin{eqnarray}\label{mesure apres avant diag}
\lefteqn{\rho_{k+1}=L(\rho_k)\mathbf{1}}
\nonumber\\
&&+\Big(-\sqrt{p(1-p)}\big(\mathcal{G}_{00}(\rho_k)+\mathcal{G}_{01}(\rho_k)\big)+\sqrt{p(1-p)}\big(\mathcal{G}_{11}(\rho_k)+\mathcal{G}_{10}(\rho_k)\big)\Big)X_{k+1}\nonumber\\&&+\sqrt{p\trace[\mathcal{G}_{01}(\rho_k)](1-p\trace[\mathcal{G}_{01}(\rho_k)])}\Bigg(-\frac{\mathcal{G}_{00}(\rho_k)}{\trace\big[\mathcal{G}_{00}(\rho_k)\big]}+\frac{\mathcal{G}_{01}(\rho_k)}{\trace\big[\mathcal{G}_{01}(\rho_k)\big]}\Bigg)Y_{k+1}^0\nonumber\\
&&+\sqrt{(1-p)\trace[\mathcal{G}_{10}(\rho_k)](1-(1-p)\trace[\mathcal{G}_{10}(\rho_k)])}\Bigg(-\frac{\mathcal{G}_{11}(\rho_k)}{\trace\big[\mathcal{G}_{11}(\rho_k)\big]}+\frac{\mathcal{G}_{10}(\rho_k)}{\trace\big[\mathcal{G}_{10}(\rho_k)\big]}\Bigg)Y_{k+1}^1.\nonumber\\
\end{eqnarray}

\begin{rk} As it was the case in equation (\ref{tata1}), the discrete random variables $X_k$ and $Y^i_k$, $i=0,1$ are centered. As before, the variables $X_k$ represent the perturbation produced by the measurement before the interaction and, given the result of this measurement, the variables $Y^i_k$ describe the perturbation generated by the measurement of the second observable. The particular choices made for $Y^i_k$ will be justified when we shall consider the continuous models. They will appear as discrete analogs of the noises which drive the continuous stochastic Master equations ($W_t$ and $\tilde{N}_t$ in equations (\ref{D}, \ref{J})).
\end{rk}

The above general framework concerns the combination of the two measurements, one before and one after each interaction. Let us present the corresponding equations when only one type of measurement (before \emph{or} after each interaction) is performed.

We start by looking at the case where a measurement is only performed before the interaction (we called this kind of setup ``Random environment''). Since measuring an observable on an element of the chain (which has not yet interacted) does not alter the state of the little system $\mathcal H$, only the reference state of each copy of $\mathcal{E}$ is random. The completely positive evolution operators describing the two possibilities for the state after the interaction are given by
\begin{eqnarray}\mathcal{R}_i(\rho)&=&\trace_\mathcal{E}[U(\rho\otimes\vert e_i\rangle\langle e_i\vert)U^\star]\\
&=&\mathcal{G}_{i0}(\rho)+\mathcal{G}_{i1}(\rho),
\end{eqnarray} for $i=0,1$. Let $\mathbf{1}^k_i$ be the random variable which is equal to $1$ if we observe the eigenvalue $\lambda_i$ at the $k$-th step, and $0$ otherwise. We can describe the evolution of the small system $\mathcal{H}$ by the following equation
$$\rho_{k+1}=\mathcal{R}_0(\rho_k)\mathbf{1}_0^{k+1}+\mathcal{R}_1(\rho_k)\mathbf{1}_1^{k+1}.$$
As before, we introduce
\[X_{k+1}=\frac{\mathbf{1}_1^{k+1}-(1-p)}{\sqrt{p(1-p)}},\,\,\,\,k\in\mathbb{N}.\]
With this notation, the evolution equation becomes
\begin{eqnarray}\label{mesure avant}
 \rho_{k+1}&=&L(\rho_k)
+\Big(-\sqrt{p(1-p)}\big(L_{00}\,\rho_k\,
L_{00}^\star+L_{10}\,\rho_k\,
L_{10}^\star\big)\nonumber\\&&\hspace{0,7cm}+\sqrt{p(1-p)}\big(L_{11}\,\rho_k\,
L_{11}^\star+L_{01}\,\rho_k\, L_{01}^\star\big)\Big)X_{k+1}.
\end{eqnarray}

The opposite case, where a measurement is only performed after the interaction, is treated in great detail in \cite{pelleg1, pelleg2} when $p=0$ (ground states) and in \cite{A1P1} for $0<p<1$ (Gibbs states). Let us recall briefly the main steps needed to obtain the appropriate equations. Consider the observable $B=\alpha_1Q_1+\alpha_2Q_2$, with $Q_i=(q_{kl}^i)_{0\leq k,l \leq 1}$. The two possible non normalized states on $\mathcal H$ that can be obtained after the measurement are defined via the action of the operators
\begin{eqnarray}
\mathcal{F}_i(\rho)&=&\trace_\mathcal{E}[I\otimes Q_i\,U(p\vert e_0\rangle\langle e_0\vert+(1-p)\vert e_1\rangle\langle e_1\vert)U^\star\,I\otimes Q_i]\\
&=&p\mathcal{G}_{0i}(\rho)+(1-p)\mathcal{G}_{1i}(\rho),
\end{eqnarray}
for $i=0,1$. The discrete evolution equation is then given by
\begin{eqnarray}
\rho_{k+1}&=&\frac{\mathcal{F}_0(\rho_k)}{\trace[\mathcal{F}_0(\rho_k)]}\mathbf{1}_0^{k+1}+\frac{\mathcal{F}_1(\rho_k)}{\trace[\mathcal{F}_1(\rho_k)]}\mathbf{1}_1^{k+1}.
\end{eqnarray}
Again, we introduce the random variables $X_k$ defined by \[X_{k+1}=\frac{\mathbf{1}_i^{k+1}-\trace[\mathcal{F}_1(\rho_k)]}{\sqrt{\trace[\mathcal{F}_0(\rho_k)]\trace[\mathcal{F}_1(\rho_k)]}}.\]
In terms of these centered random variables, we get
\begin{eqnarray}\label{sans mesure non diag}
 \rho_{k+1}&=&L(\rho_k)
\mathbf{1}
+\left(-\sqrt{\frac{\trace\big[\mathcal{F}_1(\rho_k)\big]}{\trace\big[\mathcal{F}_0(\rho_k)\big]}}\mathcal{F}_0(\rho_k)+\sqrt{\frac{\trace\big[\mathcal{F}_0(\rho_k)\big]}{\trace\big[\mathcal{F}_1(\rho_k)\big]}}\mathcal{F}_1(\rho_k)\right)X_{k+1}.
\end{eqnarray}

\section{Continuous Time Models of Quantum Trajectories}\label{sec:cont_models}
In this section, we present the continuous versions of the discrete equations (\ref{tata1}, \ref{mesure apres avant diag}, \ref{mesure avant}, \ref{sans mesure non diag}). We start by introducing asymptotic assumptions for the interaction unitaries in terms of the time parameter $\tau$. Next, we implement these assumptions in the different equations (\ref{tata1}, \ref{mesure apres avant diag}, \ref{mesure avant}, \ref{sans mesure non diag}) and we obtain stochastic differential equations as limits when the time step $\tau$ goes to $0$.
\bigskip

Let us present the asymptotic assumption for the interaction with $\tau=1/n$. In terms of the  parameter $n$ we can write the unitary operator $U$ as
\begin{equation}
 U(n)=\exp\left(-i\frac{1}{n}H_{\textrm{tot}}\right)=\left(\begin{array}{cc}
                                                  L_{00}(n)&L_{10}(n)\\
L_{01}(n)&L_{11}(n)
                                                 \end{array}
\right).
\end{equation}
Let us recall that the discrete dynamic of quantum repeated
interactions is given by
$V_k=U_k\cdots U_1.$
In \cite{AJ, AP}, it is shown that the asymptotic of the coefficients $L_{ij}(n)$ must be properly rescaled in order to obtain a non-trivial limit for $V_{[nt]}$. With proper rescaling, is it shown in these references that the operator $V_{[nt]}$ converges when $n$ goes to infinity to an operator $\tilde{V}_t$ which satisfies a \emph{Quantum Langevin Equation}. When translated in our context of a two-level atom in contact with a spin chain, we put
\begin{eqnarray}\label{assumpt}
 L_{00}(n)&=&\I+\frac{1}{n}W+\circ\left(\frac{1}{n}\right)\nonumber\\
L_{01}(n)&=&\frac{1}{\sqrt{n}}S+\circ\left(\frac{1}{\sqrt{n}}\right)\nonumber\\
L_{10}(n)&=&\frac{1}{\sqrt{n}}T+\circ\left(\frac{1}{\sqrt{n}}\right)\nonumber\\
L_{11}(n)&=&\I+\frac{1}{n}Z+\circ\left(\frac{1}{n}\right).
\end{eqnarray}
In terms of total Hamiltonian, it is shown in \cite{AP} that typical Hamiltonian $H_{\textrm{tot}}$  which gives rise to such asymptotic assumption can be described as
\[H_{\textrm{tot}}=H_{0}\otimes \I+\I\otimes
\left(\begin{array}{cc}\gamma_0&0\\0&\gamma_1\end{array}\right)+{\sqrt{n}}\Bigg(C\otimes\left(\begin{array}{cc}0&0\\1&0\end{array}\right)+
C^\star\otimes\left(\begin{array}{cc}0&1\\0&0\end{array}\right)\Bigg).\]
Hence, for the operators $W, S, T$ and $Z$ we get
\begin{eqnarray}
 W&=&-H_0-\gamma_0 \I-\frac{1}{2}C^\star C\nonumber\\
Z&=&-H_0-\gamma_1 \I-\frac{1}{2}C C^\star\nonumber\\
S&=&T^\star=-iC
\end{eqnarray}
In the rest of the paper, we shall write all the results in terms of the operators $H_0$ and $C$.

Now, we are in position to investigate the asymptotic behavior of the different equations (\ref{tata1}, \ref{mesure apres avant diag}, \ref{mesure avant}, \ref{sans mesure non diag}) and to introduce the continuous models.  The mathematical arguments used to obtain the continuous models are developed in Section 5. Before presenting the main result concerning the model with measurement, we treat the simpler model obtained by considering the limit $n$ goes to infinity in the equation (\ref{mmaster}) of Proposition \ref{reducedd}.

\subsection{Continuous Quantum Repeated Interactions without Measurement}

In this section, by applying the asymptotic assumption, we show that the limit evolution obtained from the quantum repeated interactions model is a Lindblad evolution (also called Markovian evolution \cite{F1}). This result has been stated and proved in \cite{AP}. We recall it here since the more general situations treated in the current work build upon these considerations. The discrete Master equation (\ref{mmaster}) of Proposition \ref{reducedd} in our
context is expressed as follows
\begin{equation}
 \rho_{k+1}=L(\rho_k) = p\big(L_{00}\rho_kL_{00}^\star+L_{10}\rho_kL_{10}^\star\big)+(1-p)\big(L_{01}\rho_kL_{01}^\star+L_{11}\rho_kL_{11}^\star\big).
\end{equation}
Plugging in the asymptotic assumptions (\ref{assumpt}), we get (here, $n$ is a parameter)
\begin{equation}
\rho_{k+1}=\rho_k  +\frac{1}{n}\left[p\left(-i[H_0,\rho]-\frac{1}{2}\{C^\star C,\rho\}+C\rho C^\star\right)\\ +(1-p)\left(-i[H_0,\rho]-\frac{1}{2}\{CC^\star,\rho\}+C^\star\rho C\right)+\circ(1) \right],
\end{equation}
where $[X,Y] = XY - YX$ and $\{X, Y\} = XY + YX$ are the usual commutator and anti-commutator. The following theorem is obtained by taking the limit $n \to \infty$ in the previous equation.
\begin{thm}\textbf{(Limit Model for Quantum Repeated Interactions without Measurement)}\label{DEWM}
 Let $(\rho_{[nt]})$ be the family of states defined from the sequence $(\rho_k)$ describing quantum repeated interactions. We have
$$\lim_{n\rightarrow\infty}\Vert\rho_{[nt]}-\rho_t\Vert=0,$$
where $(\rho_t)$ is the solution of the Master
equation
\[d\rho_t = \mathcal{L}(\rho_t)dt,\]
with the Lindblad operator $\mathcal{L}$ given by
\begin{eqnarray}\label{heat} \mathcal{L}(\rho)&=&p\bigg(-i[H_0,\rho]-\frac{1}{2}\{C^\star C,\rho\}+C\rho
C^\star\bigg)+(1-p)\bigg(-i[H_0,\rho]-\frac{1}{2}\{CC^\star,\rho\}+C^\star\rho
C\bigg).\nonumber\\
\end{eqnarray} 
\end{thm}

The operator $\mathcal L$ appearing in equation (\ref{heat}) is the usual Lindblad operator describing the evolution of a system in contact with a heat bath at positive temperature $T$ \cite{AJ}; let us recall that the parameter $p$ can be expressed in terms of the temperature $T$ as in equation (\ref{Gibbsstate}).

\subsection{Continuous Quantum Repeated Interactions with Measurement}

In this section, we present the different continuous models obtained as limits of discrete quantum repeated measurement models described in the equations (\ref{tata1}, \ref{mesure apres avant diag}, \ref{mesure avant}, \ref{sans mesure non diag}). 

Although continuous quantum trajectories have been extensively studied by the second author in \cite{pelleg1, pelleg2, pelleg3}, the result concerning the combination of the two kinds of measurement is new and the stochastic differential equations appearing at the limit have, to our knowledge, never been considered in the literature. The comparison between the different limiting behaviors is particularly interesting and will be discussed in detail in Section \ref{sec:discussion}.

\subsubsection{The ``Random environment'' setup}\label{RE}

In Section \ref{QMRE}, we have seen that the evolution of the little
system in presence of measurement before each interaction is described by the following equation:
\begin{eqnarray}
 \rho_{k+1}&=&L(\rho_k)+\Big(-\sqrt{p(1-p)}\big(L_{00}\,\rho_k\,
L_{00}^\star+L_{10}\,\rho_k\,
L_{10}^\star\big)\nonumber\\&&\hspace{0,7cm}+\sqrt{p(1-p)}\big(L_{11}\,\rho_k\,
L_{11}^\star+L_{01}\,\rho_k\, L_{01}^\star\big)\Big)X_{k+1}.
\end{eqnarray}
Using the asymptotic condition for the operator $L_{ij}(n)$, we get the following expression
\begin{equation}\label{as1}
 \rho_{k+1}=\rho_k+\frac{1}{n}\Big(\mathcal{L}(\rho_k)+\circ(1)\Big)+\frac{1}{n}\Big(\mathcal{K}(\rho_k)+\circ(1)\Big)X_{k+1},
\end{equation}
where the expression of $\mathcal{L}$ is the same as Theorem $1$. The accurate expression of $\mathcal{K}$ is not necessary because these terms disappears at the limit.
From the equation (\ref{as1}), we want to derive a discrete stochastic differential equation. To this end, we define the following stochastic processes
\begin{eqnarray}\label{CCF}
 \rho_n(t)&=&\rho_{[nt]},\,\,\,\,\,\,
V_n(t)=\frac{[nt]}{n},\,\,\,\,\,\,
W_n(t)=\frac{1}{\sqrt{n}}\sum_{k=0}^{[nt]-1}X_{k+1}
\end{eqnarray}
Next, by writing
$$\rho_{[nt]}=\rho_0+\sum_{k=0}^{[nt]-1}\rho_{k+1}-\rho_k$$
and by using the equation (\ref{as1}) and the definition (\ref{CCF})
of stochastic processes, we can write
\begin{equation}\label{discretesto}
 \rho_n(t)=\rho_0+\int_0^t\mathcal{L}(\rho_n(s-))dV_n(s)+\int_0^t\frac{1}{\sqrt{n}}\mathcal{E}(\rho_n(s-))dW_n(s)+\varepsilon_n(t),
\end{equation}
where $\varepsilon_n(t)$ regroups all the $\circ(\cdot)$ terms.
The equation (\ref{discretesto}) appears then as a discrete stochastic differential equation whose solution is the process $(\rho_n(t))_t$.

In order to obtain the final convergence result, we shall use the following proposition concerning the limit behavior of the process $(W_n(t))$.
\begin{pr}\label{DI}
 Let $(W_n(t))$ be the process defined by the formula $(\ref{CCF})$. We have the following convergence result
$$W_n(t)\Rightarrow W_t,$$
where $\Rightarrow$ denotes the convergence in distribution for stochastic processes and $(W_t)_{t \geq 0}$ is a standard Brownian motion.
\end{pr}
\begin{pf}
In this case, the random variables $(X_{k+1})$ are independent and identically
  distributed. Furthermore they are centered and reduced.
  As a consequence, the convergence result is just an application of the Donsker
  Theorem \cite{bilingsley, jacodshiria, protter}.
\end{pf}
\bigskip

Using Proposition \ref{DI}, we can now take the limit $n \to \infty$ in equation (\ref{discretesto}).

\begin{thm}\textbf{(Limit Model for Random Environment)}
The stochastic process $(\rho_n(t))$, describing the evolution of the small system in contact
with a random environment, converges in distribution to the
solution of the Master equation
$$d\rho_t=\mathcal{L}(\rho_t)dt,$$
where Lindblad generator $\mathcal{L}$ is given in equation (\ref{heat}).
\end{thm}

This theorem is a straightforward application of a well-known theorem of
Kurtz and Protter \cite{MR1119837, MR1112406} concerning the convergence of stochastic
differential equations. Without the term $1/\sqrt{n}$, the process $(W_n(t))$ converges to a Brownian motion and thus the equation (\ref{discretesto}) converges to a diffusive stochastic differential equation. As the term $1/\sqrt{n}$ converges to zero, this implies the random diffusive part disappears when we consider the limit. The fact that we recover the deterministic Lindblad evolution for a heat bath will be discussed in Section 3.

The next subsection contains the description of the continuous model when a measurement is performed after each interaction.

\subsubsection{Usual indirect Quantum Measurement}\label{UIQM}

In \cite{pelleg1, pelleg2}, it is shown that discrete
quantum trajectories for $p=0$ converge (when n goes to infinity) to
solutions of classical stochastic
Master equations (\ref{D}, \ref{J}). These models are at zero temperature. The
result for positive temperature ($0<p<1$) is treated in \cite{A1P1}. In this section, we just
recall the result of \cite{A1P1} corresponding to the limit models obtained from the equation (\ref{sans mesure non diag}).

As it is
mentioned in Section \ref{QMRE}, the final stochastic differential equations depend on the form of the observable.
\begin{enumerate}
 \item If $B=\alpha_0Q_0+\alpha_1Q_1$ is a diagonal observable, with $Q_i=(q_{kl}^i)_{0\leq
k,l\leq1}$, we have $q_{00}^0=q^1_{00}=1$ and all the other coefficients are equal to zero. Hence, we obtain the following asymptotic expression for the equation $(\ref{sans mesure non diag})$
\begin{eqnarray}\label{asym12}
\rho_{k+1}-\rho_k=\frac{1}{n}\Big(\mathcal{L}(\rho_{k+1})+\circ(1)\Big)+\frac{1}{n}\Big(\mathcal{N}(\rho_k)+\circ(1)\Big)X_{k+1}.
\end{eqnarray}
For the random variables $(X_k)$, we have
\begin{equation}\label{asym13}
 \left\{\begin{array}{ccc}X_{k+1}(0)=-\sqrt{\frac{\trace[\mathcal{F}_1(\rho_k)]}{\trace[\mathcal{F}_0(\rho_k)]}}&\textrm{with probability}&p+\frac{1}{n}\Big(h(\rho_k)+\circ(1)\Big)\\&&\\X_{k+1}(1)=
 \sqrt{\frac{\trace[\mathcal{F}_0(\rho_k)]}{\trace[\mathcal{F}_1(\rho_k)]}}&\textrm{with probability}&1-p+\frac{1}{n}\Big(g(\rho_k)+\circ(1)\Big)
                \end{array}\right.
\end{equation}
In (\ref{asym12}) and (\ref{asym13}), the exact expressions of $\mathcal{N}$, $h$ and $g$ are not necessary for the
final result. The expression of $\mathcal{L}$ corresponds to the Lindblad operator of Proposition 1.
\item The other case concerns an observable $B$ which is not diagonal. We have then $0<q_{00}^0<1$ and $0<q_{11}^1<1$. The final
 result is essentially the same for all non diagonal observables $B$. Hence, we just focus on the symmetric case where $B$ is of the form
\[B=\alpha_0
\begin{pmatrix}
1/2&1/2\\
1/2&1/2
\end{pmatrix}
+\alpha_1
\begin{pmatrix}
1/2&-1/2\\
-1/2&1/2
\end{pmatrix}
.\] 
Thus, in asymptotic form, the
equation $(\ref{sans mesure non diag})$ becomes
\begin{eqnarray}\label{as2}
\rho_{k+1}-\rho_k=\frac{1}{n}\Big(\mathcal{L}(\rho_k)+\circ(1)\Big)+\frac{1}{\sqrt{n}}\Big(\mathcal{G}(\rho_k)+\circ(1)\Big)X_{k+1},
\end{eqnarray}
where $\mathcal{G}$ is defined on the set of states by
\begin{eqnarray*}\mathcal{G}(\rho)&=&-\Big(p(C\rho+\rho C^\star)+(1-p)(C^\star\rho+\rho C)\Big)\\&&+\trace\Big[p(C\rho+\rho C^\star)+(1-p)(C^\star\rho+\rho C)\Big]\rho
\end{eqnarray*}
The random variables $X_{k+1}$ evolve as
 \begin{equation}
 \left\{\begin{array}{ccc}X_{k+1}(0)=-\sqrt{\frac{\trace[\mathcal{F}_1(\rho_k)]}{\trace[\mathcal{F}_0(\rho_k)]}}&\textrm{with probability}&\frac{1}{2}+\frac{1}{\sqrt{n}}\Big(f(\rho_k)\Big)\\&&\\X_{k+1}(1)=
 \sqrt{\frac{\trace[\mathcal{F}_0(\rho_k)]}{\trace[\mathcal{F}_1(\rho_k)]}}&\textrm{with probability}&\frac{1}{2}+\frac{1}{\sqrt{n}}\Big(m(\rho_k)\Big)
                \end{array}\right.
\end{equation}
Again, the exact expressions of $f$ and $m$ are not necessary for the final result.
\end{enumerate}
We define
\begin{eqnarray*}
 \rho_n(t)&=&\rho_{[nt]},\,\,\,\,\,\,
V_{[nt]}=\frac{[nt]}{n},\,\,\,\,\,\,
W_n(t)=\frac{1}{\sqrt{n}}\sum_{k=0}^{[nt]-1}X_{k+1}.
\end{eqnarray*}
Depending on which type of observable we consider, we
obtain two different discrete stochastic differential equations.
\begin{enumerate}
 \item In the case of a diagonal observable we have
\begin{eqnarray*}
 \rho_n(t)&=&\rho_0+\sum_{k=0}^{[nt]-1}\Big(\frac{1}{n}(\mathcal{L}(\rho_k)+\circ(1))+\frac{1}{\sqrt{n}}(\mathcal{N}(\rho_k)+\circ(1))\frac{1}{\sqrt{n}}X_{k+1}\Big)\\
&=&\rho_0+\int_0^t\mathcal{L}(\rho_n(s-)dV_n(s)+\int_0^t\frac{1}{\sqrt{n}}\mathcal{N}(\rho_n(s-))dW_n(s)+\varepsilon_n(t).
\end{eqnarray*}
\item In the same way, in the non diagonal case we obtain
\begin{eqnarray*}
 \rho_n(t)&=&\rho_0+\sum_{k=0}^{[nt]-1}\Big(\frac{1}{n}(\mathcal{L}(\rho_k)+\circ(1))+(\mathcal{G}(\rho_k)+\circ(1))\frac{1}{\sqrt{n}}X_{k+1}\Big)\\
&=&\rho_0+\int_0^t\mathcal{L}(\rho_n(s-)dV_n(s)+\int_0^t\mathcal{G}(\rho_n(s-))dW_n(s)+\varepsilon_n(t).
\end{eqnarray*}

\end{enumerate}
In these equations the terms $\varepsilon_n(t)$ regroup the $\circ(\cdot)$ terms. The final results are gathered in the following theorem (see \cite{A1P1} for a complete proof).

\begin{thm}\label{DEEWM}\textbf{(Limit Model for Usual Indirect Quantum Measurement)}

 Let $B$ be a diagonal observable. Let $(\rho_n(t))$ be the stochastic process defined
 from the discrete quantum trajectory describing the quantum repeated measurement of
 $B$. This stochastic process converges in distribution to the solution of the master equation
$$d\rho_t=\mathcal{L}(\rho_t)dt.$$
\smallskip

Let $B$ be a non-diagonal observable. Let $(\rho_n(t))$ be the
stochastic process defined from the discrete quantum trajectory
describing the quantum repeated measurement of $B$. This
stochastic process converges in distribution to the solution of
the stochastic differential equation
$$d\rho_t=\mathcal{L}(\rho_t)dt+\mathcal{G}(\rho_t)dW_t$$
where $(W_t)$ is a standard Brownian motion.
\end{thm}

It is important to notice that for a diagonal observable we end up with a Master equation without random terms. In \cite{pelleg2}, at zero temperature, it is shown that the limit evolution is described by a jump stochastic differential equation. Similar evolutions for diagonal observables will be recovered when we consider both measurements. The discussion in Section \ref{sec:discussion} will turn around such results.

\subsubsection{Continuous Model of Usual Indirect Quantum Measurement in Random Environment}\label{CQTRE}

This section contains the main result of the article. To our knowledge, a random environment model has never been considered before in the setup of indirect quantum measurement (neither in discrete, nor in the continuous case). 

 We treat separately the case of a diagonal observable
and a non diagonal observable. We show that for a diagonal observable, we recover a evolution including jump random times. The
limit evolution is although different as the case of \cite{pelleg2}.
\bigskip 

Let us start with the non diagonal case. As in Section \ref{UIQM}, we focus on the case
\[B=\alpha_0
\begin{pmatrix}
1/2&1/2\\
1/2&1/2
\end{pmatrix}
+\alpha_1
\begin{pmatrix}
1/2&-1/2\\
-1/2&1/2
\end{pmatrix}
.\] 
In this situation, the asymptotic form of the equation $(\ref{mesure apres avant diag})$ is given by
\begin{eqnarray}\label{asympp}
 \rho_{k+1}&=&\rho_k+\frac{1}{n}\Big(\mathcal{L}(\rho_k)+\circ(1)\Big)+\frac{1}{n}\Big(\mathcal{K}(\rho_k)+\circ(1)\Big)X_{k+1}\nonumber\\
&&+\frac{1}{\sqrt{n}}\Bigg(-\sqrt{1-(1-p)^2}\Big(C\rho_k+\rho_k C^\star-\trace\big[\rho_k(C+C^\star\big]\,\rho_k\Big)\Bigg)Y_{k+1}^0\nonumber\\
&&+\frac{1}{\sqrt{n}}\Bigg(\sqrt{(1-p^2)}\Big(C^\star\rho_k+\rho_k
C-\trace[\rho_k(C^\star+C]\rho_k\Big)\Bigg)Y_{k+1}^0
\end{eqnarray}
From the equation (\ref{asympp}), we want to derive a discrete stochastic differential equation. To this aim, we define the processes
\begin{eqnarray*}
 \rho_n(t)&=&\rho_{[nt]},\\
V_n(t)&=&\frac{[nt]}{n},\quad\quad\quad\quad\quad\quad\quad\quad
W_n(t)=\frac{1}{\sqrt{n}}\sum_{k=0}^{[nt]-1}X_{k+1},\\
W_n^0(t)&=&-\frac{1}{\sqrt{n}}\sum_{k=0}^{[nt]-1}Y^0_{k+1},\quad\quad\quad
W_n^1(t)=\frac{1}{\sqrt{n}}\sum_{k=0}^{[nt]-1}Y^1_{k+1},
\end{eqnarray*}
and the operators
\begin{eqnarray}
\mathcal{Q}(\rho)&=&\sqrt{1-(1-p)^2}\Big(C\rho+\rho C^\star-\trace[\rho(C+C^\star)]\rho\Big)\label{r1} \\
\mathcal{W}(\rho)&=&\sqrt{(1-p^2)}\Big(C^\star\rho+\rho C-\trace[\rho(C^\star+C)]\rho\Big)\label{r2}.
\end{eqnarray}
This way, the process $(\rho_n(t))$ satisfies the following discrete stochastic differential equation
\begin{eqnarray*}
 \rho_n(t)&=&\int_0^t\mathcal{L}(\rho_n(s-)dV_n(s)+\int_0^t\frac{1}{\sqrt{n}}\mathcal{K}(\rho_n(s-))dW_n(s)\\
&&+\int_0^t\mathcal{Q}(\rho_n(s-))dW_n^0(s)+\int_0^t\mathcal{W}(\rho_n(s-))dW_n^1(s)+\varepsilon_n(t).
\end{eqnarray*}
Heuristically, if we assume that
\[(W_n(t),W_n^0(t),W_n^1(t))\Longrightarrow(W_t,W_t^1,W_t^2),\]
where the processes $(W_t)$ and $(W_t^1)$ and $(W_t^2)$ are
independent Brownian motions, the following theorem becomes natural (the rigorous proof is presented in Section \ref{sec:proofs}).

\begin{thm}\label{diff}\textbf{(Limit Model for Indirect Quantum Measurement of non-diagonal observables in Random environment)}
Let $(\rho_n(t))$ be the stochastic process defined from the discrete quantum trajectory $(\rho_k)$ which describes the repeated measurement of a non-diagonal observable in random environment. Then the process $(\rho_n(t))$ converges in distribution to the solution of the stochastic differential equation
\begin{equation}\label{diff2noise}
 \rho_t=\rho_0+\int_0^t\mathcal{L}(\rho_s)ds+\int_0^t\mathcal{Q}(\rho_s)dW^1_s+\int_0^t\mathcal{W}(\rho_s)dW^2_s,
\end{equation}
where $(W_t^1)$ and $(W_t^2)$ are two independent Brownian motions.
\end{thm}

 It is important to notice that we get two
Brownian motion at the limit whereas in Theorem \ref{DEEWM} there is only one Brownian motion. We have already described a situation where the random noise disappears.
\bigskip

 Let us now deal with the diagonal case. In asymptotic form, the equation $(\ref{tata1})$ becomes
\begin{eqnarray}
 \rho_{k+1}&=&\rho_k+\frac{1}{n}\Big(\mathcal{L}(\rho_k)+\circ(1)\Big)+\frac{1}{n}\Big(\mathcal{E}(\rho_k)+\circ(1)\Big)X_{k+1}\nonumber\\
&&+\Bigg(\frac{C\rho_k C^\star}{\trace\big[C\rho_kC^\star\big]}-\rho_k+\circ(1)\Bigg)\Big(\mathbf{1}_{01}-\frac{p}{n}\big(\trace\big[C\rho_kC^\star\big]+\circ(1)\big)\Big)\nonumber\\
&&+\Bigg(\frac{C^\star\rho_k
C}{\trace\big[C^\star\rho_kC\big]}-\rho_k+\circ(1)\Bigg)\Big(\mathbf{1}_{10}-\frac{1-p}{n}\big(\trace\big[C^\star\rho_kC\big]+\circ(1)\big)\Big).
\end{eqnarray}
Such an equation can be written in the following way
\begin{eqnarray*}
\rho_{k+1}&=&\rho_k+\frac{1}{n}\Big(\mathcal{L}(\rho_k)+p\big(-C\rho_kC^\star+\trace\big[C\rho_kC^\star\big]\rho_k\big)\nonumber\\&&\hspace{1,5cm}+(1-p)\big(-C^\star\rho_kC+\trace\big[C^\star\rho_kC\big]\rho_k\big)+\circ(1)\Big)\nonumber\\
&&+\frac{1}{n}\Big(\mathcal{E}(\rho_k)+\circ(1)\Big)X_{k+1}\nonumber\\
&&+\Bigg(\frac{C\rho_k
C^\star}{\trace\big[C\rho_kC^\star\big]}-\rho_k+\circ(1)\Bigg)\mathbf{1}_{01}+\Bigg(\frac{C^\star\rho_k
C}{\trace\big[C^\star\rho_kC\big]}-\rho_k+\circ(1)\Bigg)\mathbf{1}_{10}
\end{eqnarray*}
In order to define the discrete stochastic differential equation, we need to introduce the operator
$$\mathcal{T}(\rho)=\mathcal{L}(\rho)+p\big(-C\rho C^\star+\trace\big[C\rho C^\star\big]\rho\big)+(1-p)\big(-C^\star\rho C+\trace\big[C^\star\rho C\big]\rho\big)$$
and the following processes
\begin{eqnarray}
 \rho_n(t)&=&\rho_{[nt]}\\
V_n(t)&=&\frac{[nt]}{n},\,\,\,\,\,\,\,\,\,W_n(t)=\frac{1}{\sqrt{n}}\sum_{k=0}^{[nt]-1}X_{k+1}\\
\tilde{N}_n^1(t)&=&\sum_{k=0}^{[nt]-1}\mathbf{1}_{01}\,\,\,\,\,\,\,\,\,\,\tilde{N}_n^2(t)=\sum_{k=0}^{[nt]-1}\mathbf{1}_{10}
\end{eqnarray}
We obtain a discrete stochastic differential equation
\begin{eqnarray}
 \rho_n(t)&=&\rho_0+\int_0^t\mathcal{T}(\rho_n(s-)dV_n(s)+\int_0^t\frac{1}{\sqrt{n}}\mathcal{E}(\rho_n(s-))dW_n(s)\nonumber\\
&&+\int_0^t\Bigg(\frac{C\rho_k
C^\star}{\trace[C\rho_kC^\star]}-\rho_k\Bigg)\tilde{N}_n^1(s)+\int_0^t\Bigg(\frac{C^\star\rho_k
C}{\trace[C^\star\rho_kC]}-\rho_k\Bigg)\tilde{N}_n^2(s).
\end{eqnarray}
Let us motivate briefly what follows concerning the convergence
of $(\tilde{N}_n^1(t))$ and $(\tilde{N}_n^2(t))$ (this will be rigorously justified in Section $3$).
Let us deal with $(\tilde{N}_n^1(t))$ for example. By definition
of $\mathbf{1}_{01}$, we have
\begin{equation}
 \left\{\begin{array}{ccc}
  \mathbf{1}_{01}(\omega_{k+1},\varphi_{k+1})=1&\textrm{with probability}&\frac{1}{n}\big(p\trace[C\rho_kC^\star]+\circ(1)\big)\\&&\\
\mathbf{1}_{01}(\omega_{k+1},\varphi_{k+1})=0&\textrm{with
probability}&1-\frac{1}{n}\big(p\trace[C\rho_kC^\star]+\circ(1)\big)
        \end{array}\right.
\end{equation}
Hence, for a large $n$, the random variable $\mathbf{1}_{01}$ takes the value 1 with a low probability and $0$ with a high probability. This behavior is typical of the classical Poisson process \cite{2MR704559, 2MR636252}. Heuristically we can consider a counting process $(\tilde{N}_t^1)$ as the continuous limit of $(\tilde{N}_n^1(t))$. Since a counting
process is entirely determined by its intensity (\cite{2MR636252, jacod}), we can guess its intensity by computing $\E[\tilde{N}_n^1(t)]$. We have
\begin{eqnarray}
 \E[\tilde{N}_n^1(t)]&=&\sum_{k=0}^{[nt]-1}\frac{1}{n}\E\bigg[p
 \trace[C\rho_kC^\star]+\circ(1)\bigg]\nonumber\\
&=&\int_0^t\mathbf{E}\bigg[p\trace[C\rho_n(s-)C^\star]\bigg]dV_n(s)+\tilde{\varepsilon}_n(t)
\end{eqnarray}
Assuming that the processes $(\tilde{N}_n^1(t))$ and
$(\rho_n(t))$ converge, we get
\[\E[\tilde{N}_t^1]=\int_0^t\mathbf{E}\bigg[p\trace[C\rho_{s-}C^\star]\bigg]ds.\]
We thus define the limit process $(\tilde{N}_t^1)$ as
a counting process with stochastic intensity
$t\rightarrow\int_0^tp\trace[C\rho_{s-}C^\star]ds$. In the same way, we assume that $(\tilde{N}_n^2(t))$ converges to a counting process $(\tilde{N}_t^2)$
with stochastic intensity 
$t\rightarrow\int_0^t(1-p)\trace[C^\star\rho_{s-}C]ds$.

 The limit stochastic differential equation would then be 
\begin{eqnarray}
d\rho_t=\mathcal{T}(\rho_{t-})dt+\Bigg(\frac{C\rho_{t-}C^\star}{\trace[C\rho_{t-}C^\star]}-\rho_{t-}\Bigg)d\tilde{N}_t^1+\Bigg(\frac{C^\star\rho_{t-}C}{\trace[C^\star\rho_{t-}C]}-\rho_{t-}\Bigg)d\tilde{N}_t^2.
\end{eqnarray}
From a mathematical point of view, the way of defining this equation is not absolutely rigorous because the definition of the driving processes depends on the solution $(\rho_t)$ (usually, in order to define solutions of a stochastic differential equation, one needs to consider previously the driving processes).

A rigorous way to introduce this equation consists in defining it in terms of two
Poisson Point processes $N^1$ and $N^2$ on $\mathbb{R}^2$ which are mutually
independent (see \cite{pelleg2, jacodprotter}). More precisely, we consider the stochastic
differential equation
\begin{eqnarray}\label{GGOO}
 \rho_t&=&\rho_0+\int_0^t\mathcal{T}(\rho_{s-})ds+\int_0^t\int_{\mathbb{R}}\Bigg(\frac{C\rho_{s-}C^\star}{\trace\big[C\rho_{s-}C^\star\big]}-\rho_{s-}\Bigg)\mathbf{1}_{0<x<p\trace[C\rho_{s-}C^\star]}N^1(ds,dx)\nonumber\\
&&+\int_0^t\int_{\mathbb{R}}\Bigg(\frac{C^\star\rho_{s-}C}{\trace\big[C^\star\rho_{s-}C\big]}-\rho_{s-}\Bigg)\mathbf{1}_{0<x<(1-p)\trace[C^\star\rho_{s-}C]}N^2(ds,dx)
\end{eqnarray}
This allows to write the equation in an intrinsic way and, if (\ref{GGOO}) admits a solution, we can define the processes
 \begin{eqnarray}
  \tilde{N}_t^1&=&\int_0^t\int_{\mathbb{R}}\mathbf{1}_{0<x<p\trace[C\rho_{s-}C^\star]}N^1(ds,dx)\,\,\,\,\textrm{and}\nonumber\\
\tilde{N}_t^2&=&\int_0^t\int_{\mathbb{R}}\mathbf{1}_{0<x<(1-p)\trace[C^\star\rho_{s-}C]}N^2(ds,dx).
 \end{eqnarray}

We can now state the convergence theorem in this context.

\begin{thm}\label{poisson}\textbf{(Limit Model for Indirect Quantum Measurement of diagonal observables in Random environment)}

 Let $N^1$ and $N^2$ be two independent Poisson point processes on $\mathbb{R}^2$ defined on a probability space $(\Omega,\mathcal{F},\P)$. Let $(\rho_n(t))$ be the process defined from the discrete quantum trajectory $(\rho_k)$ which describes the measurement of a diagonal observable $A$ in a random environment. The stochastic process $(\rho_n(t))$ converges in distribution to the solution of the stochastic differential equation
\begin{eqnarray}\label{poiss2noise}
 \rho_t&=&\rho_0+\int_0^t\mathcal{T}(\rho_{s-})ds+\int_0^t\int_{\mathbb{R}}\Bigg(\frac{C\rho_{s-}C^\star}{\trace\big[C\rho_{s-}C^\star\big]}-\rho_{s-}\Bigg)\mathbf{1}_{0<x<p\trace[C\rho_{s-}C^\star]}N^1(ds,dx)\nonumber\\
&&+\int_0^t\int_{\mathbb{R}}\Bigg(\frac{C^\star\rho_{s-}C}{\trace\big[C^\star\rho_{s-}C\big]}-\rho_{s-}\Bigg)\mathbf{1}_{0<x<(1-p)\trace[C^\star\rho_{s-}C]}N^2(ds,dx)
\end{eqnarray}
\end{thm}

\begin{rk}
It is not obvious that the
stochastic differential equations (\ref{diff}, \ref{poisson}) admit a unique solution (even in the diffusive case, the coefficients are not Lipschitz and the jump term can vanish). The uniqueness questions is treated in \cite{pelleg1, pelleg2, jacodprotter, jacod}.
\end{rk}

Let us stress at this point that in this article we have focused on the particular case $\mathcal{E}=\mathbb{C}^2$. This case allows to consider observables with two different eigenvalues. In \cite{pelleg3,A1P1}, situations with more than two eigenvalues are considered but only when measurements are performed after the interactions. The statistical model (Random environment) is not treated. In this article, our aim was to compare the situation with and without measurement before the interaction in order to emphasize the situations appearing in the case of random environment. The situation that we have treated is sufficiently insightful to point out the differences between the statistical model and the Gibbs model. Higher dimension can easily be treated by adapting the presentation of this article and the results of \cite{pelleg3,A1P1}; the continuous evolutions involve mixing between jump and diffusion evolution (see also \cite{Book,infinite,B6} for other references on such types of equations).

In the following section, we compare the different continuous stochastic Master equations in the different model of environment.

\section{Discussion}\label{sec:discussion}

The different models we have considered and the limiting continuous equations that govern the dynamics are summed up in Table \ref{tab:B_measurement}. Each cell of the table contains the type of evolution equation in the zero temperature case ($T=0$) and in the positive temperature case ($T>0$). Hence, in what follows, the parameter $p$, until now supposed constant, will be allowed to vary. Continuous, Master equations evolutions are denoted by $\mathcal L_p$ where $p$ is the parameter related to the temperature ($T=0$ corresponds to $p=1$). In these terms the two differential equation at $T=0$ are given by
\begin{equation}\label{rr}
d\rho_t=\mathcal{L}_1(\rho_{t-})dt+\left(\frac{C\rho_{t-}C^\star}{\trace\big[C\rho_{t-}C^\star\big]}-\rho_{t-}\right)\Big(d\tilde{N}_t-\trace\big[C\rho_{t-}C^\star\big]dt\Big),
\end{equation}
where $(\tilde{N}_t)$ is a counting process with stochastic intensity $\int_0^t\trace\big[C\rho_{s-}C^\star\big]ds$ and
\begin{equation}\label{rrd}
d\rho_t=\mathcal{L}_1(\rho_{t-})dt+\Big(C\rho_t+\rho_tC^\star-\trace\Big[\rho_t(C+C^\star)\Big]\,\rho_t\Big)dW_t,
\end{equation}
where $(W_t)$ is a Brownian motion.

\begin{table}
\caption{Different models and the corresponding continuous behavior}
\bigskip
\begin{tabular}{|c|c|c|c|c|}
\hline
& No measurement & Before & After & Before \& After \\
\hline
$B$ diagonal & \multirow{2}{*}{\begin{tabular}{c}$T=0 \quad \mathcal L_1$ \\ $T>0 \quad \mathcal L_p$ \end{tabular}} & \multirow{2}{*}{\begin{tabular}{c} $T=0 \quad \mathcal L_1$ \\ $T>0 \quad \mathcal L_p$ \end{tabular}} & \begin{tabular}{c} $T=0 \quad 1J$ \\ $T>0 \quad \mathcal L_p$ \end{tabular} & \begin{tabular}{c} $T=0 \quad 1J$ \\ $T>0 \quad 2J$ \end{tabular}  \\
\cline{1-1} \cline{4-5} 
$B$ non-diagonal & & & \begin{tabular}{c} $T=0 \quad 1D$ \\ $T>0 \quad 1D$ \end{tabular} & \begin{tabular}{c} $T=0 \quad 1D$ \\ $T>0 \quad 2D$ \end{tabular} \\
\hline
\end{tabular}
\label{tab:B_measurement}
\end{table}

Note that when no measurement is performed after the interaction (the ``No measurement'' and ``Before'' columns), the type of the observable $B$ is irrelevant. Moreover, at zero temperature, the measurement before the interaction is irrelevant, since the state of the system to be measured is an eigenstate of the observable. Hence the last two columns contain identical information in the case $T=0$. 

The discussion that follows is meant to provide insight about this table and on the different limit behaviors that appear. We shall try, as much as possible, to provide physical explanations for the similarities and differences between the different models treated in the present work. 

\subsection{Gibbs vs. Statistical models at near zero temperatures}

In order to emphasize the differences between the statistical model and the Gibbs model, we investigate the stochastic equations when the parameter $p$ goes to 1, that is the temperature goes to zero (this fact is related to the assumption $\gamma_0>\gamma_1$ in the description of the free Hamiltonian of $\mathcal{E}$). In particular, we show that we can recover the zero temperature case from the statistical model by considering the limit p goes to $1$, while it is not the case in the Gibbs model. This can be seen in the case of a diagonal observable. At zero temperature for a diagonal observable, the continuous model is given by the jump equation (\ref{rr}). In the Gibbs model, for a diagonal observable, we get only the master equation $d\rho_t=\mathcal{L}_p(\rho_t)dt$. It is then obvious that we do not recover the equation (\ref{rr}) when we consider the limit $p$ goes to one. Concerning the statistical model, i.e random environment, the limit equation is given by
\begin{eqnarray}\label{rrr}
d\rho_t&=&\mathcal{L}_p(\rho_{t-})dt+\left(\frac{C\rho_{t-}C^\star}{\trace\big[C\rho_{t-}C^\star\big]}-\rho_{t-}\right)\Big(d\tilde{N}^1_t-p\trace\big[C\rho_{t-}C^\star\big]dt\Big)\nonumber\\&&+\left(\frac{C^\star\rho_{t-}C}{\trace\big[C^\star\rho_{t-}C\big]}-\rho_{t-}\right)\Big(d\tilde{N}^2_t-(1-p)\trace\big[C^\star\rho_{t-}C\big]dt\Big),
\end{eqnarray}
where $(\tilde{N}^1_t)$ is a counting process with stochastic intensity $\int_0^tp\trace\big[C\rho_{s-}C^\star\big]ds$ and $(\tilde{N}^2_t)$ is a counting process with stochastic intensity $\int_0^t(1-p)\trace\big[C^\star\rho_{s-}C\big]ds$. Heuristically, if we consider the limit $p=1$, we get a counting process $(\tilde{N}^2_t)$ with a intensity equal to zero and $(\tilde{N}^1_t)$ is a counting process with stochastic intensity $\int_0^t\trace\big[C\rho_{s-}C^\star\big]ds$. As a consequence, we have that almost surely, for all $t$ $\tilde{N}^2_t=0$. Hence, we recover the equation (\ref{rr}) at the limit $p=1$ (this result can be rigorously proved by considering the limit $p=1$ in the Markov generator see Section \ref{sec:proofs}). Let us notice that the limit $p=1$ in the diffusive evolution allows to recover the model at zero temperature for the diffusive evolution in both models (statistical and Gibbs).

\subsection{Gibbs vs. Statistical Models: absorption and emission interpretation}

In the preceding section, we have seen that the Gibbs model and the Statistical model give rather different continuous evolution equations, especially in the case where a diagonal observable is measured. We are now going to provide a more complete interpretation of the Table \ref{tab:B_measurement}. To this end, we shall concentrate on the special case where $\mathcal{H}=\mathcal{E}$ and 
\[C = \begin{pmatrix}
	0 & 0 \\
	1 & 0
\end{pmatrix}.\]
This particular choice for the Hamiltonians is known as the \emph{dipole type interaction model} and it has the property that the interaction between the small system and each copy of the chain is symmetric. This will allow us to give an interpretation of the evolution of the small system in terms of emisions and absorption of photons. In such a setup, we shall clearly identify and explain the differences between the two models (Gibbs and Statistical).

Let us start by commenting on the similarities between these models. If no measure is performed after each interaction, we have seen that the limit evolution is the same in both models. In particular, the randomness generated by the measure in the Statistical model disappears at the limit and we get a classical Master equation. 

The models become different when one considers a measurement, after each interaction. As in the previous section, the differences are more significant in the case where the measured observable $B$ is diagonal. In order to illustrate the differences between the two models, we start by describing the trajectory of the solutions of the jump equations and by explaining the apparition of jumps. 

At zero temperature, the evolution equation (\ref{rr}) can be re-written as
\begin{equation}\label{rrrr}
d\rho_t=\mathcal{S}_1(\rho_{t-})dt+\left(\frac{C\rho_{t-}C^\star}{\trace\big[C\rho_{t-}C^\star\big]}-\rho_{t-}\right)d\tilde{N}_t
\end{equation}
by regrouping the $dt$ terms. The solution of such a stochastic differential equation can be described in following manner. Let $(T_n)_n$ the jump times of the counting process $(\tilde{N}_t)$, that is $T_n=\inf\{t/\tilde{N}_t=n\}$. We have then
\begin{equation}
\rho_t=\int_0^t\mathcal{S}_1(\rho_{s-})ds+\sum_{k=0}^\infty\left(\frac{C\rho_{T_k-}C^\star}{\trace\big[C\rho_{T_k-}C^\star\big]}-\rho_{T_k-}\right)\mathbf{1}_{T_k\leq t}.
\end{equation}
This expression is rigorously justified in \cite{pelleg2}. What this means is that in the time intervals between the jumps, the solution satisfies the ordinary differential equation $d\rho_t=\mathcal{S}_1(\rho_{t-})dt$ and at jump times its discontinuity is given by
\begin{equation}
\rho_{T_k}=\rho_{T_k-}+\frac{C\rho_{T_k-}C^\star}{\trace\big[C\rho_{T_k-}C^\star\big]}-\rho_{T_k-}=\frac{C\rho_{T_k-}C^\star}{\trace\big[C\rho_{T_k-}C^\star\big]}.
\end{equation}

In a similar fashion, the solution of equation (\ref{rrr}) satisfies
\begin{eqnarray*}
\rho_t&=&\int_0^t\mathcal{S}_p(\rho_{s-})ds+\sum_{k=0}^\infty\left(\frac{C\rho_{T^1_k-}C^\star}{\trace\big[C\rho_{T^1_k-}C^\star\big]}-\rho_{T^1_k-}\right)\mathbf{1}_{T^1_k\leq t}\\&&+\sum_{k=0}^\infty\left(\frac{C^\star\rho_{T^2_k-}C}{\trace\big[C^\star\rho_{T^2_k-}C\big]}-\rho_{T^2_k-}\right)\mathbf{1}_{T^2_k\leq t}
\end{eqnarray*}
where for $i=0,1$, the terms $(T^i_k)$ correspond to the jump times of the processes $(\tilde{N}_t^i)$. Depending on the type of the jump, the discontinuity of the solution is given by
\begin{equation}
\rho_{T_k^1}=\frac{C\rho_{T^1_k-}C^\star}{\trace\big[C\rho_{T^1_k-}C^\star\big]}\,\,\textrm{or}\,\,\rho_{T_k^2}=\frac{C^\star\rho_{T^2_k-}C}{\trace\big[C^\star\rho_{T^2_k-}C\big]}.
\end{equation}
\begin{rk} Since the two Poisson point processes $N^1$ and $N^2$ are independent, on the probability space $(\Omega,\mathcal{F},\P)$ supporting these two processes, we have
\[\P\left[\{\omega\in\Omega\; |\; \exists k\in\mathbb{N}, T_k^1(\omega)=T_k^2(\omega)\}\right]=0.\]
This means that a jump of type $1$ cannot occur at the same time as a jump of type $2$. We shall see later on that this condition is also physically relevant.
\end{rk}
An explicit computation with the particular value of $C$ we considered gives
\begin{equation}
\frac{C\rho C^\star}{\trace\big[C\rho C^\star\big]}=\left(\begin{array}{cc}1&0\\0&0\end{array}\right)=\vert e_0\rangle\langle e_0\vert\,\,\textrm{and}\,\,\frac{C^\star\rho C}{\trace\big[C^\star\rho C\big]}=\left(\begin{array}{cc}0&0\\0&1\end{array}\right)=\vert e_1\rangle\langle e_1\vert,
\end{equation}
for all states $\rho$.

In the setup with two possible jumps, depending on the type of jump, the state of the small system after the jump is either ground state or the excited state. This has a clear interpretation in terms of the emission and absorption of photons. At zero temperature, it is well known that equation (\ref{rrrr}) describes an counting photon experiment \cite{B5,BAR} and that the a jump corresponds to the emission of a photon, which will be detected by the measuring apparatus. In the case where two jumps can occur, the same interpretation remains valid for type $1$ jumps (emission of a photon). After such an emission, the state of the small system is projected on the ground state $\ketbra{e_0}{e_0}$. Type $2$ jumps are characterized by the fact that the state of the small systems jumps to the excited state $\ketbra{e_1}{e_1}$; this corresponds to the absorption of a photon by the small system, which justifies its excitation. Note that the impossibility of simultaneous jumps of the two types (see the above remark) is physically justified by the fact that the small system can not absorb and emit a photon in the same time.

\begin{figure}
	\centering
		\includegraphics{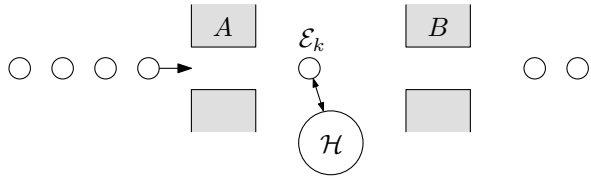}
	\caption{Experimental setup}
	\label{fig:setup}
\end{figure}

This interpretation has a clear meaning in the discrete model. Let us consider the experimental setup in Figure \ref{fig:setup}. This setup, with two measuring apparatus, corresponds to the Statistical model.

\begin{itemize}
\item At zero temperature, each copy of $\mathcal E$ is in the ground state $\ketbra{e_0}{e_0}$. In this case, the first measurement device will never click and only the result of the second apparatus is relevant. If, at the step $k+1$, the second apparatus does not click, the state of the small system is given by $\rho_{k+1}=L_{00}\rho_kL_{00}^\star/\trace[L_{00}\rho_kL_{00}^\star]$. In the asymptotic regime, we get $\rho_{k+1}=\rho_k+1/n\mathcal{S}_1(\rho_k)+\circ(1)$, which is an approximation of a continuous evolution. On the other hand, if the second apparatus clicks, then the evolution is given by $\rho_{k+1}=L_{10}\rho_kL_{10}^\star/\trace[L_{10}\rho_kL_{10}^\star]=C\rho_kC^\star/\trace[C\rho_k C^\star]+\circ(1)$, which corresponds to the emission of a photon. This corresponds to a jump, as indicated by the result of the measurement.
\item At positive temperature, both devices can click. If the first apparatus does not click, the state of $\mathcal E$ before the interaction is $\ketbra{e_0}{e_0}$ and we have the same interpretation of the second measurement as before. In the other case, a click for the first measurement implies that the state of $\mathcal E$ is $\ketbra{e_1}{e_1}$. Now, the interpretation of the second measurement is the following. If we have a click, then the evolution is continuous, and the absence of a click corresponds to a jump of the form $C^\star\rho_k C/\trace[C^\star\rho_k C]+\circ(1)$ (absorption of a photon). Let us stress that this corresponds to the inverse of the situation where no click occurs at the first measurement. The different cases are summarized in the Table \ref{tab:physical_interpretation}.

\begin{table}
\caption{Physical interpretation of measurements}
\begin{tabular}{|c|c|c|}
\hline
App. $A$ $\setminus$ \;  App. $B$ & No click & Click \\
\hline
No click & Continuous & Emission \\
\hline
Click & Absorption & Continuous \\
\hline
\end{tabular}
\label{tab:physical_interpretation}
\end{table}
\end{itemize}

We are now in the position to explain the difference between the Gibbs and the Statistical models. In the Statistical model, the first measurement allows us to clearly identify if the small system absorbs or emits a photon. If we consider the same experiment without the first measurement device, we obtain the Gibbs model. In this setup, the information provided by the second apparatus is not sufficient to distinguish between a continuous evolution, an absorption or an emission. Indeed, as it has been pointed out in the above description, in order to have the exact variations of the state of the small system, it is necessary to know if the state of $\mathcal E$ is $\ketbra{e_0}{e_0}$ or $\ketbra{e_1}{e_1}$ \emph{before} the interaction.

\subsection{Unraveling}

In order to conclude the Section \ref{sec:discussion}, we shall to investigate an important physical feature called \emph{unraveling}. This concept is related with the possibility to describe the stochastic master equations in terms of pure states. More precisely, an important category of stochastic master equations preserve the property of being valued in the set of pure states, that is if the initial state is pure, then, at all times, the state of the small system will continue to be pure. This property is of great importance for numerical simulations; indeed, less parameters are needed to describe a pure state than an arbitrarily density matrix (for a $K$-dimensional Hilbert space, a pure state is "equivalent" to a vector that is we need $2K-1$ real parameters, whereas for a density matrix we need $K^2$ such real coordinates). Since the expectation of the solution of a stochastic master equation reproduces the solution of the master equation, by taking the average of a large number of simulations of the stochastic master equations we get a simulation of the master equation. An important gain of simulation is obtained by the pure state property. This technique is called \emph{Monte Carlo Wave Function Method}. 

When a stochastic Master equation preserve the property of being a pure state, it is said that the stochastic master equations gives an unraveling of the master equation (or unravels the master equation). In this setup, one can express a stochastic differential equation for vectors in the underlying Hilbert space. This equation is called \emph{stochastic Schr\"odinger equation}. In this subsection, we want to show that the continuous models obtained from the limit of the repeated measurements before and after the interaction give rise to unraveling of the master equation for a heat bath whereas the unraveling property is not satisfied if we consider the measurement only after the interaction. Let us stress that at zero temperature, this property has already been established in \cite{pelleg1,pelleg2} (the author do not refer to unraveling but he shows that the stochastic master equations (\ref{rr}) and (\ref{rrd}) preserve the property of being valued in the pure states set).

In order to obtain the expression of the stochastic Schr\"odinger equation for the heat bath, we show that the quantum trajectories can be expressed in terms of pure states. To this end, we show that for all $k$, there exists a norm $1$ vector $\psi_k\in\mathcal{H}$ such that $\rho_k=\ketbra{\psi_k}{\psi_k}$. Next, by considering the process $\psi_t^{(n)}$ and the convergence when $n$ goes to infinity, we get a stochastic differential equation for norm one vectors in $\mathcal{H}$. Following the form of observables, we obtain two types of equations, which are equivalent of (\ref{rr}) and (\ref{rrd}) (the equivalence is characterized by the fact that a solution $(\psi_t)$ of an equation for vectors allow to consider the process $(\ketbra{\psi_t}{\psi_t})$ which satisfies the corresponding stochastic master equation).

We proceed by recursion. Let suppose that there exists $\psi_k$ such that $\rho_k=\ketbra{\psi_k}{\psi_k}$. Let $Q_i$ be one of the eigenprojectors of the observable $B$ which is measured after the interaction. Since $Q_i$ is a one dimensional projector, there exists a norm $1$ vector $\vartheta_i$ such that $Q_i=\ketbra{\vartheta_i}{\vartheta_i}$. For $j\in\{0,1\}$, the transitions between $\rho_{k+1}$ and $\rho_k$ are given by the non normalized operators $\xi_{k+1}(ji)=\trace_{\mathcal{E}}\big[I\otimes Q_i\,\,U(\rho_k\otimes\ketbra{e_j}{e_j})U^\star\,\,I\otimes Q_i\big]$, for $(i,j)\in\{0,1\}^2$ and we have
\begin{eqnarray}
\xi_{k+1}(ji)&=&\trace_{\mathcal{E}}\big[I\otimes \ketbra{\vartheta_i}{\vartheta_i}\,\,U(\ketbra{\psi_k}{\psi_k}\otimes\ketbra{e_j}{e_j})U^\star\,\,I\otimes \ketbra{\vartheta_i}{\vartheta_i}\big]\nonumber\\
&=&\trace_{\mathcal{E}}\left[I\otimes \ketbra{\vartheta_i}{\vartheta_i}\,\,\sum_{p,l}L_{pl}\ketbra{e_p}{e_l}\big(\ketbra{\psi_k}{\psi_k}\otimes\ketbra{e_j}{e_j}\big)\sum_{u,v}L_{uv}^\star\ketbra{e_v}{e_u}\,\,I\otimes \ketbra{\vartheta_i}{\vartheta_i}\right]\nonumber\\
&=&\trace_{\mathcal{E}}\left[\sum_{p,l,u,v}L_{pl}\ketbra{\psi_k}{\psi_k}L_{uv}^\star\,\,\otimes\,\,\ketbra{\vartheta_i}{\vartheta_i}\ketbra{e_p}{e_l}\ketbra{e_j}{e_j}\ketbra{e_v}{e_u}\ketbra{\vartheta_i}{\vartheta_i} \right]\nonumber\\
&=&\trace_{\mathcal{E}}\left[I\otimes \ketbra{\vartheta_i}{\vartheta_i}\,\,\sum_{p,l}L_{pl}\ketbra{e_p}{e_l}\big(\ketbra{\psi_k}{\psi_k}\otimes\ketbra{e_j}{e_j}\big)\sum_{u,v}L_{uv}^\star\ketbra{e_v}{e_u}\,\,I\otimes \ketbra{\vartheta_i}{\vartheta_i}\right]\nonumber\\
&=&\trace_{\mathcal{E}}\left[\sum_{p,u}L_{pj}\ketbra{\psi_k}{\psi_k}L_{uj}^\star\,\,\otimes\,\,\ketbra{\vartheta_i}{\vartheta_i}\ketbra{e_p}{e_u}\ketbra{\vartheta_i}{\vartheta_i} \right]\nonumber\\
&=&\trace_\mathcal{E}\left[\left\vert\sum_p\langle e_p;\vartheta_i\rangle L_{pj}\,\psi_k\right\rangle\left\langle\sum_u\langle e_u;\vartheta_i\rangle L_{uj}\,\psi_k\right\vert\otimes\ketbra{\vartheta_i}{\vartheta_i}\right]\nonumber\\
&=&\left\vert\sum_p\langle e_p;\vartheta_i\rangle L_{pj}\,\psi_k\right\rangle\left\langle\sum_u\langle e_u;\vartheta_i\rangle L_{uj}\,\psi_k\right\vert.
\end{eqnarray}
We can then define \begin{equation}\label{rree}\psi_{k+1}(ji)=\frac{\sum_p\langle e_p;\vartheta_i\rangle L_{pj}\,\psi_k}{\Vert\sum_p\langle e_p;\vartheta_i\rangle L_{pj}\,\psi_k\Vert}\mathbf{1}_{ji},\end{equation}
which describe the evolution of the wave function of $\mathcal{H}$. This equation is equivalent to the discrete stochastic Master equations in the sense that almost surely (with respect to $\P$) $\ketbra{\psi_k}{\psi_k}=\rho_k$, for all $k$. Let us stress that, here, the normalizing factor $\Vert\sum_p\langle e_p;\vartheta_i\rangle L_{pj}\,\psi_k\Vert$ appearing in the quotient is not the probability of outcome. Indeed, the probability of outcome is $\Vert\sum_p\langle e_p;\vartheta_i\rangle L_{pj}\,\psi_k\Vert^2$. Now, we can investigate the continuous limit of this equation by applying the asymptotic assumptions described in Section \ref{sec:cont_models}. Depending on the form of the observable $B$, we obtain two different kinds of equations:
\begin{itemize}
\item A jump equation ($B$ diagonal)
\begin{eqnarray}\label{Ret}
\psi_t&=&\psi_0+\int_0^t\mathcal{F}_p(\psi_{s-})ds+\int_0^t\int_\mathbb{R}\left(\frac{C\psi_{s-}}{\sqrt{\mu_{s-}}}-\psi_{s-}\right)\mathbf{1}_{0<x<p\mu_{s-}}N^1(dx,ds)\nonumber\\
&&+\int_0^t\int_\mathbb{R}\left(\frac{C^\star\psi_{s-}}{\sqrt{\nu_{s-}}}-\psi_{s-}\right)\mathbf{1}_{0<x<(1-p)\nu_{s-}}N^2(dx,ds),
\end{eqnarray}
where \begin{eqnarray}\mathcal{F}_p(\psi_{s-})&=&p\left(-iH_0-\frac{1}{2}\big(C^\star C+\mu_{s-}I\big)+\sqrt{\mu_{s-}}C\right)\psi_{s-}\nonumber\\&&+(1-p)\left(-iH_0-\frac{1}{2}\big(C C^\star+\nu_{s-}I\big)+\sqrt{\nu_{s-}}C^\star\right)\psi_{s-}\end{eqnarray}
and $\mu_{s-}=\langle \psi_{s-},C^\star C\psi_{s-}\rangle$ and $\nu_{s-}=\langle \psi_{s-},C C^\star\psi_{s-}\rangle.$\item A diffusive equation ($B$ non diagonal)
\begin{eqnarray}\label{Ret1}
\psi_t&=&\psi_0+\int_0^t\mathcal{G}_p(\psi_s)ds+\int_0^t\sqrt{1-(1-p^2)}(C-\kappa_sI)\psi_sdW_s^1\nonumber\\&&+\int_0^t\sqrt{(1-p^2)}(C^\star-\zeta_sI)\psi_sdW_s^2,
\end{eqnarray}
where
\begin{eqnarray}
\mathcal{G}_p(\psi_s)&=&p\left(-iH_0-\frac{1}{2}\big(C^\star C-2\kappa_sC+\kappa_s^2I\big)\right)\psi_s\nonumber\\&&+(1-p)\left(-iH_0-\frac{1}{2}\big(C C^\star-2\zeta_sC^\star+\zeta_s^2I\big)\right)\psi_s,
\end{eqnarray}
with $\kappa_s=\textrm{Re}(\langle\psi_s,C\psi_s\rangle)$ and $\zeta_s=\textrm{Re}(\langle\psi_s,C^\star\psi_s\rangle)$.
\end{itemize}  
By applying the It\^{o} rules in stochastic calculus, we can make the following observation which establishes the connection between the equations for vectors and the equations for states. Let $(\psi_t)$ be the solution of equation (\ref{Ret}) (respectively (\ref{Ret1})), then almost surely $\ketbra{\psi_t}{\psi_t}=\rho_t$, for all $t\geq 0$, where $(\rho_t)$ is the solution of (\ref{poiss2noise}) in Theorem \ref{poisson} (respectively (\ref{diff2noise}) in Theorem \ref{diff}). Such considerations are the continuous equivalent of the remark following equation (\ref{rree}).

In other words, the description of the evolution of the system $\mathcal{H}$ in the setup with both measurements can be described in terms of pure states. A key property for unraveling is that
\begin{equation}
d\mathbf{E}\big[\ketbra{\psi_t}{\psi_t}\big]=\mathcal{L}_p\big(\E\big[\ketbra{\psi_t}{\psi_t}\big]\big)dt,
\end{equation}
for any solutions of (\ref{Ret}) or (\ref{Ret1}).

\section{Proofs of Theorems $\ref{diff}$ and $\ref{poisson}$}\label{sec:proofs}

The last section of the paper is devoted to showing that discrete quantum
trajectories in random environment converge to solutions of stochastic differential equations (\ref{diff2noise}, \ref{poiss2noise}). We proceed in the following way.

In a first step, we justify rigorously the form of the stochastic
differential equations provided in Theorems $\ref{diff}$ and $\ref{poisson}$. Starting with the description of discrete quantum trajectories in terms of Markov chains, we can define the so called \emph{discrete Markov generators} of these Markov chains. These generators depend naturally on the parameter $n$ of the length of interaction. When $n$ goes to infinity, the limit of the discrete Markov generators gives rise to infinitesimal generators. Next, these limit generators can be naturally associated with \emph{problems of martingale} \cite{jacod, kurtz}. The solution of martingale
problems associated with these generators (see Definition \ref{def:pb_mg} below) can be then expressed in terms of solutions of particular stochastic differential equations.  We show that the appropriate equations are the same as the ones in Theorems $\ref{diff}$ and $\ref{poisson}$. This justifies the heuristic presentation of (\ref{diff2noise}, \ref{poiss2noise}) in Section \ref{CQTRE}. 

This first step provides actually the convergence of finite dimensional laws of the discrete quantum trajectories to the continuous one. Finally, in a second step, we prove the total convergence in distribution by showing that the discrete quantum trajectories own the property of \textit{tightness} (see \cite{bilingsley, jacodshiria}). 

\subsection{Convergence of Markov Generators and Martingale problems}

Let us start by defining the infinitesimal generator of a discrete quantum trajectory. Let $B=\alpha_0 Q_0+\alpha_1 Q_1$ be an observable where $Q_i=(q_{kl}^i)$. Let $(\rho_k)$ be any quantum trajectory describing the measurement of the observable $A$ in a random environment with initial state $\rho_0$. Using the Markov property (Proposition \ref{mmaa} in Section \ref{DEV}) of $(\rho_k)$ on $(\Omega,\mathcal{C},\P)$, we can consider the process
  $(\rho_n(t))$ which satisfies
  \begin{eqnarray}
\P[\rho_n(0)=\rho]&=&1\nonumber\\
\P\left[\rho_n(t)=\rho_k\; \Big| \;\frac{k}{n}\leq t<\frac{k+1}{n}\right]&=&1\,\,\,\,\textrm{for all}\,\,k\nonumber\\
\P\big[\rho_{k+1}\in B \; | \; \rho_k=\rho\big]&=&\Pi(\rho,B)\,\,\,\,\textrm{for all Borel sets}\,\,B,
  \end{eqnarray}
where $\Pi$ is the transition function of the Markov chain
$(\rho_k)$. More precisely, the transition
function $\Pi$ is defined, for all Borel sets $B$, by
\begin{eqnarray*}\Pi(\rho,B)&=&p\trace[\mathcal{G}_{00}(\rho)]\delta_{h_0(\rho)}(B)+p\trace[\mathcal{G}_{01}(\rho)]
\delta_{h_1(\rho)}(B)\\&&+(1-p)\trace[\mathcal{G}_{10}(\rho)]\delta_{g_0(\rho)}(B)+(1-p)\trace[\mathcal{G}_{11}(\rho)]\delta_{g_1(\rho)}(B),
\end{eqnarray*}
where, for $i=0,1$, we recall that
\begin{eqnarray}\mathcal{G}_{0i}(\rho)&=&q^i_{00}L_{00}\rho L_{00}^\star+q^i_{10}L_{00}\rho L_{10}^\star+q^i_{01}L_{10}\rho L_{00}^\star+q^i_{11}L_{10}\rho L_{10}^\star\nonumber\\
\mathcal{G}_{1i}(\rho)&=&q^i_{00}L_{01}\rho
L_{01}^\star+q^i_{10}L_{01}\rho L_{11}^\star+q^i_{01}L_{11}\rho
L_{01}^\star+q^i_{11}L_{11}\rho L_{11}^\star\nonumber\\
h_i(\rho)&=&\frac{\mathcal{G}_{0i}(\rho)}{\trace[\mathcal{G}_{0i}(\rho)]},\,\,\,\,\,\,\,\textrm {and}\,\,\,\,\,\,\,
g_i(\rho)=\frac{\mathcal{G}_{1i}(\rho)}{\trace[\mathcal{G}_{1i}(\rho)]}.
\end{eqnarray}
It is worth noticing that the transition function $\Pi$ is
defined on the set of states. The discrete Markov generator of the
Markov process $(\rho_n(t))$ is defined as
\begin{equation*}\mathcal{A}_nf(\rho)=n\int_{E}\Big(f(\mu)-f(\rho)\Big)\Pi(\rho,d\mu),\end{equation*}
where $E$ denotes the set of states and $f$ is any function of class $\mathcal C^2$
with compact support. The set of such functions is denoted
by $\mathcal C^2_c(E)$. In our situation, for all $f\in \mathcal C_c^2(E)$, we have
 \begin{eqnarray}\label{defgenerattor}
 \mathcal{A}_nf(\rho)&=&n\Bigg(p\trace[\mathcal{G}_{00}(\rho)]\Big(f(h_0(\rho))-f(\rho)\Big)+p\trace[\mathcal{G}_{01}(\rho)]
 \Big(f(h_1(\rho))-f(\rho)\Big)\nonumber\\
 &&(1-p)\trace[\mathcal{G}_{10}(\rho)]\Big(f(g_0(\rho))-f(\rho)\Big)+(1-p)\trace[\mathcal{G}_{11}(\rho)]
 \Big(f(g_1(\rho))-f(\rho)\Big)\Bigg).\nonumber\\
 \end{eqnarray}
 Now, we can implement the asymptotic assumptions (\ref{assumpt}) introduced at the beginning of Section \ref{sec:cont_models} and we can consider the limit of $\mathcal{A}_n$ when $n$ goes to infinity. In a similar way as Section \ref{CQTRE}, the result is divided into two parts depending on the form of the observable $B$. 

 \begin{pr}\label{convgenn}Let $\mathcal{A}_n$ be the infinitesimal generator of
 the discrete quantum trajectory describing the measurement of a
 diagonal observable. We have for all $f\in \mathcal C^2_c(E)$
 \begin{eqnarray}
\lim_{n\rightarrow\infty}\sup_{\rho\in
E}\Vert\mathcal{A}_nf(\rho)-\mathcal{A}^jf(\rho)\Vert=0,
 \end{eqnarray}
where $\mathcal{A}^j$ is an infinitesimal generator defined, for all $f\in \mathcal C^2_c(E)$, by
\begin{eqnarray}\mathcal{A}^jf(\rho)&=&D_\rho f\big(\mathcal{L}(\rho)\big)+\left[f\left(\frac{C\rho C^\star}{\mathrm{Tr}[C\rho C^\star]}\right)-f(\rho)-D_\rho f\left(\frac{C\rho C^\star}{\mathrm{Tr}[C\rho C^\star]}-\rho\right)\right]\mathrm{Tr}\big[C\rho C^\star\big]\nonumber\\&&+\left[f\left(\frac{ C^\star\rho C}{\mathrm{Tr}[ C^\star\rho C]}\right)-f(\rho)-D_\rho f\left(\frac{ C^\star\rho C}{\mathrm{Tr}[ C^\star\rho C]}-\rho\right)\right]\mathrm{Tr}\big[C^\star\rho C\big].
\end{eqnarray}

Let $\mathcal{A}_n$ be the infinitesimal generator of
 the discrete quantum trajectory describing the measurement of the
 non-diagonal observable
\[B=\alpha_0
\begin{pmatrix}
1/2&1/2\\
1/2&1/2
\end{pmatrix}
+\alpha_1
\begin{pmatrix}
1/2&-1/2\\
-1/2&1/2
\end{pmatrix}
.\] 
We have for all $f\in \mathcal C^2_c(E)$
 \begin{eqnarray}
\lim_{n\rightarrow\infty}\sup_{\rho\in
E}\Vert\mathcal{A}_nf(\rho)-\mathcal{A}^df(\rho)\Vert=0,
 \end{eqnarray}
where $\mathcal{A}^d$ is an infinitesimal generator defined, for all $f\in \mathcal C^2_c(E)$, by
$$\mathcal{A}^df(\rho)=D_\rho f\big(\mathcal{L}(\rho)\big)+\frac{1}{2}D^2_\rho f(\mathcal{Q}(\rho),\mathcal{Q}(\rho))+\frac{1}{2}D^2_\rho f(\mathcal{W}(\rho),\mathcal{W}(\rho)),$$
where $\mathcal{Q}$ and $\mathcal{W}$ are defined by the expressions $(\ref{r1})$ and $(\ref{r2})$. 
 \end{pr}

We do not provide the proof of this proposition (similar computations are presented in great detail in \cite{pelleg3}). Now, we can introduce the martingale problem associated with the limit generators of the above Proposition \ref{convgenn}. To this aim, we denote $(\mathcal{F}^\mu_t)$, the filtration generated by a process $(\mu_t)$, where $\mathcal{F}^\mu_t=\sigma(\mu_s,s\leq t)$ for all $t\geq0$.

\begin{df}\label{def:pb_mg}
Let $(\Omega,\mathcal{F},\P)$ be a probability space. Let
$i\in\{j,d\}$ and let $\rho_0$ be a state on $E$. A solution
associated with the problem of martingale $(\mathcal{A}^i,\rho_0)$
is a process $(\rho^i_t)$ such that, for all $f\in \mathcal C^2_c$, the
process $(M_t^i(f))$ defined by
$$M_t^i(f)=f(\rho^i_t)-f(\rho_0)-\int_0^t\mathcal{A}^if(\rho_s^i)ds$$
is a $(\mathcal{F}_t^{\rho^i})$ martingale with respect to $\P$.
\end{df}
Usually, solutions of stochastic differential equations are used to solve the problems of martingale \cite{jacod, kurtz}. In our context, we recover the stochastic differential equations (\ref{diff2noise}, \ref{poiss2noise}) introduced in Theorems $\ref{diff}$ and $\ref{poisson}$. Let us start by the non diagonal case.

\begin{thm}\textbf{(Solution of the Problem of Martingale for a Non-Diagonal Observable)}
Let $(\Omega,\mathcal{F},\P)$ be a probability space which supports
two independent Brownian motions $(W_t^1)$ and $(W_t^2)$. Let
$\mathcal{A}^d$ the infinitesimal generators corresponding to the
discrete quantum trajectory describing the measurement of
\[A=\lambda_0
\begin{pmatrix}
1/2&1/2\\
1/2&1/2
\end{pmatrix}
+\lambda_1
\begin{pmatrix}
1/2&-1/2\\
-1/2&1/2
\end{pmatrix}
.\]
Let $(\rho_0)$ be any state.  The solution of the problem of martingale associated
to $(\mathcal{A}^d,\rho_0)$ is given by the solution of the
following stochastic differential equation
\begin{equation}\label{diffusivee}
 \rho_t=\rho_0+\int_0^t\mathcal{L}(\rho_s)ds+\int_0^t\mathcal{Q}(\rho_s)dW^1_s+\int_0^t\mathcal{W}(\rho_s)dW^2_s.
\end{equation}
\end{thm}

 The equivalent theorem in the diagonal observable case is expressed as follows.

\begin{thm}\textbf{(Solution of the Problem of Martingale for a Diagonal Observable)}
Let $(\Omega,\mathcal{F},\P)$ be a probability space which supports
two independent Poisson Point Process $N^1$ and $N^2$. Let
$\mathcal{A}^j$ the infinitesimal generators corresponding to the
discrete quantum trajectory describing the measurement of a
diagonal observable. Let $\rho_0$ be any state.  The solution of the problem of martingale associated
to $(\mathcal{A}^j,\rho_0)$ is given by the solution of the
following stochastic differential equation
\begin{eqnarray}\label{ppoisson}
 \rho_t&=&\rho_0+\int_0^t\mathcal{T}(\rho_{s-})ds+\int_0^t\int_{\mathbb{R}}\Bigg(\frac{C\rho_{s-}C^\star}{\trace\big[C\rho_{s-}C^\star\big]}-\rho_{s-}\Bigg)\mathbf{1}_{0<x<p\trace[C\rho_{s-}C^\star]}N^1(ds,dx)\nonumber\\
&&+\int_0^t\int_{\mathbb{R}}\Bigg(\frac{C^\star\rho_{s-}C}{\trace\big[C^\star\rho_{s-}C\big]}-\rho_{s-}\Bigg)\mathbf{1}_{0<x<(1-p)\trace[C^\star\rho_{s-}C]}N^2(ds,dx).
\end{eqnarray}
\end{thm}
These theorems can be proved by using It\^o stochastic calculus (see \cite{pelleg3} for explicit computations).
\bigskip

In order to complete the study of the limit infinitesimal generators, we express a uniqueness theorem of solutions for the problems of martingale. Moreover this result is essential to prove the final convergence theorem.

\begin{pr}\label{UNIQ}
 Let $(\rho_0)$ be a state and let $\mathcal{A}^i$, $i=0,1$ be a generator defined in Proposition $5$. The problem of martingale $(\mathcal{A}^i,\rho_0)$ admits a unique solution in distribution. It means that two solutions of the martingale problem $(\mathcal{A}^i,\rho_0)$ have the same law.
\end{pr}

This proposition is actually a consequence of the uniqueness of
solution for the stochastic differential equation associated with
$\mathcal{A}^i$. Complete reference about Markov generators and
problems of martingales (uniqueness, existence) can be found in
\cite{kurtz}.

The next section contains the final convergence result.

\subsection{Tightness Property and Convergence Result}

We prove that discrete quantum trajectories $(\rho_n(t))$ have the tightness property (also called relative compactness for stochastic processes). Next, we show that the convergence result of Markov generators (Proposition \ref{convgenn}) implies the convergence of finite dimensional laws. The tightness property and the finite dimensional laws convergence imply then the convergence in distribution for stochastic processes \cite{bilingsley}.

Concerning the tightness property, we have the following result.

\begin{pr}\label{pr:tight}\textbf{(Tightness)}
 Let $(\rho_n(t))$ be any quantum trajectory describing the repeated quantum measurement of an observable $A$ (diagonal or not). There exists some constant $Z$ such that for all $t_1<t<t_2$
\begin{equation}\label{TTTTTT}\E\Big[\Vert\rho_n(t_2)-\rho_n(t)\Vert^2\Vert\rho_n(t)-\rho_n(t_1)\Vert^2\Big]\leq Z(t_1-t_2)^2.\end{equation}
As a consequence, the sequence of discrete processes $(\rho_n(t))$
is tight.
\end{pr}

In order to see that the property (\ref{TTTTTT}) implies the tightness property, the reader can consult \cite{bilingsley}. Before to prove the Proposition \ref{pr:tight}, we need the following Lemma.

\begin{lem}Let $(\rho_k)$ be the Markov chain describing the discrete quantum trajectory defined by the repeated quantum measurement of an observable $A$. Let
$$\mathcal{M}_r^{(n)}=\sigma\{\rho_j,j\leq r\},$$
 and let $(r,l)\in\mathbb{N}^2$ such that $r<l$. Then there exists a constant $K_A$ such that
$$\mathbf{E}\Big[\Vert\rho_l-\rho_r\Vert^2/\mathcal{M}_r^{(n)}\Big]\leq K_A\,\times\,\frac{l-r}{n}.$$
\end{lem}
\begin{pf}
 We just treat the case where $B$ is diagonal (similar reasoning yield the non diagonal case). Let us start with the term defined by $\mathbf{E}\left[\Vert\rho_l-\rho_r\Vert^2/\mathcal{M}^{(n)}_{l-1}\right]$. We have
\begin{eqnarray}\label{cdf}
 \mathbf{E}\left[\Vert\rho_l-\rho_r\Vert^2/\mathcal{M}^{(n)}_{l-1}\right]
 &=&\mathbf{E}\left[\left\Vert\sum_{i}\phi_i(\rho_{l-1}){\mathbf{1}}^{l+1}_{0i}+\sum_{i}\theta_i(\rho_{l-1}){\mathbf{1}}^{l+1}_{1i}-\rho_r
 \right\Vert^2\bigg{/}\mathcal{M}^{(n)}_{l-1}\right]\nonumber\\
&=&\mathbf{E}\left[\sum_{i}\left\Vert \phi_i(\rho_{l-1})
-\rho_r\right\Vert^2p\trace[\mathcal{G}_{0i}(\rho_{l-1})]\bigg{/}\mathcal{M}^{(n)}_{l-1}\right]\nonumber\\
&&+\mathbf{E}\left[\sum_{i}\left\Vert \theta_i(\rho_{l-1})
-\rho_r\right\Vert^2(1-p)\trace[\mathcal{G}_{1i}(\rho_{l-1})]\bigg{/}\mathcal{M}^{(n)}_{l-1}\right].
\end{eqnarray}
With the asymptotic description
of $\phi_i$ and $\theta_i$, we have for the first term in the right side of expression (\ref{cdf})
\begin{eqnarray*}
 &&\mathbf{E}\left[\sum_{i}\left\Vert \phi_i(\rho_{l-1})
-\rho_r\right\Vert^2p\trace[\mathcal{G}_{0i}(\rho_{l-1})]\bigg{/}\mathcal{M}^{(n)}_{l-1}\right]\\
&=&\mathbf{E}\left[\Vert\rho_{l-1}+\frac{1}{n}\left(\mathcal{L}_0(\rho_{l-1})+\circ(1))\right)-\rho_r\Vert^2p(1-\frac{1}{n}(\trace[C\rho_{l-1}C^\star]+\circ(1)))\bigg{/}\mathcal{M}^{(n)}_{l-1}\right]\nonumber\\
&&+\mathbf{E}\left[\Vert f_2(\rho_{l-1})-\rho_r\Vert^2\frac{p}{n}(\trace[C\rho_{l-1}C^\star]+\circ(1)))\bigg{/}\mathcal{M}^{(n)}_{l-1}\right]\nonumber\\
&\leq&p\mathbf{E}\left[\Vert\rho_{l-1}-\rho_r\Vert^2\bigg{/}\mathcal{M}^{(n)}_{l-1}\right]\nonumber\\&&+\frac{1}{n}\times\mathbf{E}\left[\Vert\frac{1}{n}\mathcal{L}_0(\rho_{l-1})+\circ(1))\Vert^2p(1-\frac{1}{n}(\trace[C\rho_{l-1}C^\star]+\circ(1))\bigg{/}\mathcal{M}^{(n)}_{l-1}\right]\nonumber\\
&&+\frac{1}{n}\times\mathbf{E}\left[\Vert
f_2(\rho_{l-1})-\rho_r\Vert^2(\trace[C\rho_{l-1}C^\star]+\circ(1)))\bigg{/}\mathcal{M}^{(n)}_{l-1}\right].
\end{eqnarray*}
As the discrete quantum trajectory $(\rho_k)$ takes values in the
set of states which is compact and as the function defined on the
set of state $\rho\longmapsto f_2(\rho)\big(\trace[C\rho C^\star]+\circ(1)\big)$
is continuous, there exists a constant $Z_1$ such that, almost
surely
\begin{eqnarray}
 \mathbf{E}\left[\sum_{i}\left\Vert \phi_i(\rho_{l-1})
-\rho_r\right\Vert^2p\trace[\mathcal{G}_{0i}(\rho_{l-1})]\bigg{/}\mathcal{M}^{(n)}_{l-1}\right]
\leq p\mathbf{E}\left[\Vert\rho_{l-1}-\rho_r\Vert^2\bigg{/}\mathcal{M}^{(n)}_{l-1}\right]+\frac{Z_1}{n}.
\end{eqnarray}
In the same way there exists a constant $Z_2$ such that
\begin{eqnarray}
\mathbf{E}\left[\sum_{i}\left\Vert \theta_i(\rho_{l-1})
-\rho_r\right\Vert^2p\trace[\mathcal{G}_{1i}(\rho_{l-1})]\bigg{/}\mathcal{M}^{(n)}_{l-1}\right]
\leq(1-p)\mathbf{E}\left[\Vert\rho_{l-1}-\rho_r\Vert^2\bigg{/}\mathcal{M}^{(n)}_{l-1}\right]+\frac{Z_2}{n}.\nonumber\\
\end{eqnarray}
Finally, for an appropriate constant $Z$, we have almost surely
\begin{eqnarray}
 \mathbf{E}\left[\Vert\rho_{l}-\rho_r\Vert^2\bigg{/}\mathcal{M}^{(n)}_{l-1}\right]\leq\mathbf{E}\left[\Vert\rho_{l-1}-\rho_r\Vert^2\bigg{/}\mathcal{M}^{(n)}_{l-1}\right]+\frac{Z}{n}.
\end{eqnarray}
As a consequence, by remarking that
\begin{eqnarray*}
\mathbf{E}\left[\Vert\rho_{l}-\rho_r\Vert^2\bigg{/}\mathcal{M}^{(n)}_{r}\right]=\E\left[\E\left[\Vert\rho_{l}-\rho_r\Vert^2\bigg{/}\mathcal{M}^{(n)}_{l-1}\right]\bigg{/}\mathcal{M}^{(n)}_{r}\right]
\end{eqnarray*}
by induction, we have
\begin{eqnarray*}
\E\left[\Vert\rho_{l}-\rho_r\Vert^2\bigg{/}\mathcal{M}^{(n)}_{r}\right]\leq
K_A\frac{l-r}{n}.
\end{eqnarray*}
In the non-diagonal case, the computation and estimation are similar and the Lemma holds.
\end{pf}
\bigskip

Proposition \ref{pr:tight} follows from this lemma. 

\begin{pf} \textbf{(Proposition \ref{pr:tight})} Thanks to Lemma $1$, for all quantum trajectories $(\rho_n(t))$, we have:
\begin{eqnarray*}
 &&\E\left[\Vert\rho_n(t_2)-\rho_n(t)\Vert^2\Vert\rho_n(t)-\rho_n(t_1)\Vert^2\right]\\
&=&\E\left[\E\left[\Vert\rho_n([nt_2])-\rho_n([nt])\Vert^2/\mathcal{M}_{[nt]}^{(n)}\right]\Vert\rho_n([nt])-\rho_n([nt_1])\Vert^2\right]\\
&\leq&\frac{K_A([nt_2]-[nt])}{n}\E\left[\E\left[\Vert\rho_n([nt])-\rho_n([nt_1])\Vert^2/\mathcal{M}_{[nt_1]}^n\right]\right]\\
&\leq&\frac{K_A([nt_2]-[nt])}{n}\frac{K_A([nt]-[nt_1])}{n}\\
&\leq&Z_A(t_2-t_1)^2,
\end{eqnarray*}
with $Z_A=4(K_A)^2$ and the result follows.
\end{pf}
\bigskip

Since the tightness property holds, it remains to prove that the finite dimensional laws converge. This result follows from the following proposition.

\begin{pr}\label{pr:conv_fin_dim_law}
 Let $\rho_0$ be a state. Let $(\rho_n(t))$ be a quantum trajectory describing a repeated quantum measurement of an observable $A$. Let $\mathcal{A}^i$, $i=0,1$ be the associated Markov generator, we have
\begin{eqnarray}
 \lim_{n\rightarrow\infty}\E\left[\left(f(\rho_n(t+s))-f(\rho_n(t)
-\int_t^{t+s}\mathcal{A}^if(\rho_n(s))\right)\prod_{i=1}^m\theta_i(\rho_n(t_i))\right]
&=&0
\end{eqnarray}
for all $m\geq0$, for all $0\leq t_1<t_2<\ldots<t_m\leq t<t+s$,
for all functions $(\theta_i)_{i=1,\ldots, m}$ and for all $f$ in $\mathcal C^2_c$.
\end{pr}
\begin{pf}
 Let $(\rho_n(t))$ be any discrete quantum trajectory and $\mathcal{A}^i$ the associated generator. Let $(\mathcal{F}_t^{(n)})$ denote the natural filtration of the process $(\rho_n(t))$, that is
\[(\mathcal{F}_t^{(n)})=\sigma\{\rho_n(s),s\leq t\}=\mathcal{M}_{[nt]}^{(n)}.\]
For $m\geq0$,  $0\leq t_1<t_2<\ldots<t_m\leq t<t+s$ and  $f, \theta_1, \ldots, \theta_m \in \mathcal C^2_c$, we have
\begin{eqnarray}
 &&\E\left[\left(f(\rho_n(t+s))-f(\rho_n(t)-\int_t^{t+s}\mathcal{A}^if(\rho_n(s))\right)\prod_{i=1}^m\theta_i(\rho_n(t_i))\right]\nonumber\\
&=&\E\left[\E\left[\left(f(\rho_n(t+s))
-f(\rho_n(t)-\int_t^{t+s}\mathcal{A}^if(\rho_n(s))\right)/\mathcal{F}_t^n\right]\prod_{i=1}^m\theta_i(\rho_n(t_i))\right].
\end{eqnarray}
Let us now estimate the term $\E\left[\left(f(\rho_n(t+s))-f(\rho_n(t)-\int_t^{t+s}\mathcal{A}^if(\rho_n(s))\right)/\mathcal{F}_t^n\right]$. To this end, from the definition of infinitesimal generators, we can notice that the discrete process defined for all $n$ by
\begin{equation}\label{martingaleproperty}
 f(\rho_n(k/n))-f(\rho_0)-\sum_{j=0}^{k-1}\frac{1}{n}\mathcal{A}^i_nf(\rho_n(j/n))
\end{equation}
is a $(\mathcal{F}_{k/n}^n)$-martingale (this is the discrete
equivalent of solutions for problems of martingale for discrete
processes).

Now, assuming $r/n\leq t<(r+1)/n$ and $l/n\leq t+s<(l+1)/n$, we have
$\mathcal{F}_t^n=\mathcal{F}_{r/n}^n$. The random states
$\rho_n(t)$ and $\rho_n(t+s)$ satisfy then
$\rho_n(t)=\rho_n(r/n)$ and $\rho_n(t+s)=\rho_n(l/n)$. The
martingale property $(\ref{martingaleproperty})$ implies then
\begin{eqnarray}
 &&\E\left[f(\rho_n(t+s))-f(\rho_n(t)\Big{/}\mathcal{F}_t^n\right]\nonumber\\
 &=&\E\left[f(\rho_n(l/n))-f(\rho_n(k/n)\Big{/}\mathcal{F}_{r/n}^n\right]\nonumber\\
&=&\E\left[\sum_{j=k}^{l-1}\frac{1}{n}\mathcal{A}^i_nf(\rho_n(j/n))\Big{/}\mathcal{F}_{r/n}^n\right]\nonumber\\
&=&\E\left[\int_t^{t+s}\mathcal{A}_n^if(\rho_n(s))ds\Big{/}\mathcal{F}_t^n\right]\nonumber\\&&
+\E\left[\left(t-\frac{r}{n}\right)\mathcal{A}_n^if(\rho_n(t))+\left(\frac{l}{n}-(t+s)\right)\mathcal{A}^i_nf(\rho_n(t+s))\Big{/}\mathcal{F}_t^n\right].
\end{eqnarray}
As a consequence, we have
\begin{eqnarray}
 &&\Bigg{\vert}\E\left[\left(f(\rho_n(t+s))-f(\rho_n(t)-\int_t^{t+s}
 \mathcal{A}^if(\rho_n(s))\right)\prod_{i=1}^m\theta_i(\rho_n(t_i))\right]
 \Bigg{\vert}\nonumber\\
&\leq&\E\left[\bigg{\vert}\int_t^{t+s}\mathcal{A}^i_nf(\rho_n(s))-
\mathcal{A}^if(\rho_n(s)ds \bigg{\vert}\right]\prod_{i=1}^m\Vert \theta_i\Vert_{\infty}\nonumber\\
&&+\E\left[\bigg{\vert}\left(t-\frac{[nt]}{n}\right)\mathcal{A}_nf(\rho_n(t))+
\left(\frac{[n(t+s)]}{n}-(t+s)\right)\mathcal{A}^i_nf(\rho_n(t+s))
\bigg{\vert}\right]\prod_{i=1}^m
\Vert \theta_i\Vert_{\infty}\nonumber\\
&\leq&
M\sup_{\rho\in\mathcal{S}}\bigg{\vert}\mathcal{A}_n^if(\rho)-\mathcal{A}^if(\rho)\bigg{\vert}+\frac{L}{n}\sup_{\rho\in\mathcal{S}}\bigg{\vert}\mathcal{A}_n^if(\rho)\bigg{\vert},
\end{eqnarray}
where $M$ and $L$ are constants depending on $\Vert h_i\Vert$ and $s$. Thanks to the condition of uniform convergence from Proposition $2$, we obtain
\begin{equation}
 \lim_{n\rightarrow\infty}\Bigg{\vert}\E\left[\left(f(\rho_n(t+s))
 -f(\rho_n(t)-\int_t^{t+s}\mathcal{A}_if(\rho_n(s))\right)\prod_{i=1}^m\theta_i(\rho_n(t_i))\right]\Bigg{\vert}=0.
\end{equation}
\end{pf}

We finish by showing that Propositions \ref{pr:tight} and \ref{pr:conv_fin_dim_law} implies the convergence in distribution.
 Indeed the tightness property, which is equivalent to relative compactness for the Topology of Skorohod \cite{jacodshiria, bilingsley}, implies that all converging subsequence of $(\rho_n(t))$ converges in distibution to the solution of the problem martingale $(\mathcal{A}^i,\rho_0)$. In other terms, let $(Y_t)$ be a limit process of a subsequence of $(\rho_n(t))$, Proposition \ref{pr:conv_fin_dim_law} implies that
\begin{eqnarray}
 \mathbf{E}\left[\left(f(Y_{t+s})-f(Y_t)
-\int_t^{t+s}\mathcal{A}^if(Y_s)ds\right)\prod_{i=1}^m\theta_i(Y_{t_i})\right]
&=&0,
\end{eqnarray}
for all $m\geq0$, for all $0\leq t_1<t_2<\ldots<t_m\leq t<t+s$,
for all functions $(\theta_i)_{i=1,\ldots, m}$ and for all $f$ in
$\mathcal C^2_c$. As a consequence $(Y_t)$ is a Markov process
(with respect to its natural filtration $(\mathcal{F}_t^Y)$),
which is also a solution of the martingale problem
$(\mathcal{A}^i,\rho_0)$. Now, the uniqueness of the solution of
the problem of martingale (Proposition \ref{UNIQ}) allows to conclude that the discrete quantum
trajectory converges in distribution to the solution of the
problem of martingale.


\begin{thebibliography}{100}

\bibitem{A1}
Attal, St\'ephane {\it Quantum noises.} Open quantum systems. II, 79--147, Lecture Notes in Math., 1881, Springer, Berlin, 2006.

\bibitem{A2}
Attal, S {\it Quantum Noises} Book to appear

\bibitem{AJ}
Attal, St\'ephane; Joye, Alain {\it The Langevin equation for a quantum heat bath.} J. Funct. Anal. 247 (2007), no. 2, 253--288.

\bibitem{AP}
Attal, St\'ephane; Pautrat, Yan {\it From repeated to continuous quantum interactions.} Ann. Henri Poincar\'e 7 (2006), no. 1, 59--104.

\bibitem{AP2}
Attal, St\'ephane; Pautrat, Yan {\it From $(n+1)$-level atom chains to $n$-dimensional noises.} Ann. Inst. H. Poincar\'e Probab. Statist. 41  
(2005), no. 3, 391--407.

\bibitem{A1P1}
Attal S and Pellegrini C
 {\it Stochastic Master Equations for a Heat Bath}.
 preprint, 2007.
 
   \bibitem{BAR}
  A.~Barchielli
  \newblock Direct and heterodyne detection and other applications of quantum stochastic calculus to quantum optics.
  \newblock Quantum Opt. 2 (1990) 423--441.

\bibitem{Book}
A.~Barchielli and M.~Gregoratti.
\newblock
Quantum Trajectories and Measurements in Continuous Time
The Diffusive Case.
\newblock {\em Lecture Notes in Physics} , Vol. 782 


\bibitem{B1}
Barchielli, Alberto {\it Continual measurements in quantum mechanics and quantum stochastic calculus.} Open quantum systems. III,  
207--292, Lecture Notes in Math., 1882, Springer, Berlin, 2006.

\bibitem{B2}
Barchielli, Alberto {\it Quantum stochastic calculus, measurements continuous in time, and heterodyne detection in quantum optics. Classical and quantum  
systems} (Goslar, 1991), 488--491, World Sci. Publ., River Edge, NJ, 1993.

\bibitem{B3}
Barchielli, Alberto; Lupieri, Giancarlo {\it Instruments and mutual entropies in quantum information.} Quantum probability, 65--80, Banach Center Publ., 73,  
Polish Acad. Sci., Warsaw, 2006.

\bibitem{B4}
Barchielli, A.; Lupieri, G. {\it Instrumental processes, entropies, information in quantum continual measurements.} Quantum Inf. Comput.  
4 (2004), no. 6-7, 437--449.

\bibitem{B5}
Barchielli, Alberto; Zucca, Fabio {\it On a class of stochastic differential equations used in quantum optics.} Rend. Sem. Mat. Fis. Milano 66  
(1996), 355--376 (1998).

\bibitem{B6}
Barchielli, A.; Paganoni, A. M.; Zucca, F. {\it On stochastic differential equations and semigroups of probability operators in quantum  
probability.} Stochastic Process. Appl. 73 (1998), no. 1, 69--86.

\bibitem{B7}
Barchielli, A.; Holevo, A. S. { \it Constructing quantum measurement processes via classical stochastic calculus.} Stochastic Process. Appl. 58  
(1995), no. 2, 293--317. 

\bibitem{Bel1}
Belavkin, Viacheslav P. {\it Quantum stochastic calculus and quantum nonlinear filtering.} J. Multivariate Anal. 42 (1992), no. 2, 171--201.

\bibitem{Bout1}
Bouten, Luc; Guta, Madalin; Maassen, Hans {\it Stochastic Schr\"odinger equations.} J. Phys. A 37 (2004), no. 9, 3189--3209.

\bibitem{Bout2}
Luc Bouten, Ramon van Handel and Matthew James: 
{\it A discrete invitation to quantum filtering and feedback control} 
To appear: SIAM Review, arXiv:math.PR/0606118

\bibitem{Bout3}
Luc Bouten, Ramon van Handel and Matthew James: 
{\it An introduction to quantum filtering} 
SIAM J. Control Optim. Vol. 46, pp. 2199-2241, 2007

\bibitem{Fagn1}
Fagnola, Franco {\it Quantum stochastic differential equations and dilation of completely positive semigroups.} Open quantum systems. II,  
183--220, Lecture Notes in Math., 1881, Springer, Berlin, 2006.

\bibitem{gardiner}
 Gardiner, C. W.; Zoller, P. {\it Quantum noise. A handbook of Markovian and non-Markovian quantum stochastic methods with applications to quantum optics.} Third edition. Springer Series in Synergetics. Springer-Verlag, Berlin, 2004.
 
 \bibitem{infinite}
C.~M.Mora and R.~Rebolledo.
\newblock Basic Properties of Non-linear Stochastic
Schr\"odinger Equations Driven by Brownian MOotions.
\newblock{\em Annals of Applied Probability} 2008, Vol. 18, No. 2, 591–619

\bibitem{NP}
Nechita, I. and Pellegrini, C. {\it Random repeated quantum interactions and random invariant states.} Preprint available at \url{http://arxiv.org/abs/0902.2634}.

\bibitem{Part1}
Parthasarathy, K. R. {\it An introduction to quantum stochastic calculus.} Monographs in Mathematics, 85. BirkhŠuser Verlag, Basel, 1992. xii 
+290 pp. ISBN: 3-7643-2697-2

\bibitem{pelleg1}
Pellegrini, C. {\it Existence, Uniqueness and Approximation of Stochastic Schr\"odinger Equation: the diffusive
case.} The Annals of Probability
2008, Vol. 36, No. 6, 2332Ð2353

\bibitem{pelleg2}
Pellegrini, C. {\it Existence, uniqueness and approximation for stochastic Schr\"odinger equation: the Poisson
case.} Preprint available at http://arxiv.org/abs/0709.3713.

\bibitem{pelleg3}
Pellegrini, C.
 {\it Markov Chains Approximation of Jump-Diffusion Quantum
  Trajectories}.
 preprint, 2008.

\bibitem{F1}
Breuer, Heinz-Peter; Petruccione, Francesco
{\it The theory of open quantum systems.} Oxford University Press, New York, 2002

\bibitem{bilingsley}
P. Billingsley. {\it Convergence of probability measures.} Wiley Series in Probability and 
Statistics : Probability and Statistics. John Wiley and Sons Inc., New York, second 
edition, 1999. A Wiley-Interscience Publication. 

\bibitem{kurtz}
 S. N. Ethier and T. G. Kurtz. {\it Markov processes.} Wiley Series in Probability and 
Mathematical Statistics : Probability and Mathematical Statistics. John Wiley and Sons 
Inc., New York, 1986. Characterization and convergence. 

\bibitem{jacod}
J. Jacod. {\it Calcul stochastique et probl\`emes de martingales,} volume 714 of Lecture Notes 
in Mathematics. Springer, Berlin, 1979. 

\bibitem{jacodprotter}
J. Jacod and P. Protter. {\it Quelques remarques sur un nouveau type dÕ\'equations 
di?\'erentielles stochastiques.} In Seminar on Probability, XVI, volume 920 of Lecture 
Notes in Math., pages 447Ð458. Springer, Berlin, 1982. 

\bibitem{jacodshiria}
J. Jacod and A. N. Shiryaev. {\it Limit theorems for stochastic processes,} volume 288 of 
Grund lehren der Mathematischen Wissenschaften [Fundamental Principles of Mathematical Sciences]. Springer-Verlag, Berlin, second edition, 2003.

\bibitem{protter}
P. E. Protter. {\it Stochastic integration and di?erential equations,}volume 21 of Applications of Mathematics (New York). Springer-Verlag, Berlin, second edition, 2004. 
Stochastic Modelling and Applied Probability.

\bibitem{MR1112406}
T.~G. Kurtz and P.~Protter.
\newblock Weak limit theorems for stochastic integrals and stochastic
  differential equations.
\newblock {\em Ann. Probab.}, 19(3):1035--1070, 1991.

\bibitem{MR1119837}
T.~G. Kurtz and P.~Protter.
\newblock Wong-{Z}akai corrections, random evolutions, and simulation schemes
  for {SDE}s.
\newblock In {\em Stochastic analysis}, pages 331--346. Academic Press, Boston,
  MA, 1991.
  
  \bibitem{2MR636252}
P.~Br{\'e}maud.
\newblock {\em Point processes and queues}.
\newblock Springer-Verlag, New York, 1981.
\newblock Martingale dynamics, Springer Series in Statistics.

\bibitem{2MR704559}
T.~C. Brown.
\newblock Some {P}oisson approximations using compensators.
\newblock {\em Ann. Probab.}, 11(3):726--744, 1983.

\bibitem{W1}
H.~M. Wiseman and G.~J Milburn
\newblock interpretation of quantum jump and diffusion processes illustrated on the Bloch sphere
\newblock {\it Phys Rev A} vol. 47.3 1652-1666 (1993)

\bibitem{W2}
H.~M. Wiseman
\newblock Quantum trajectories and feedback
\newblock Ph.D Thesis 1994
 

\end{thebibliography}
\end{document}